\documentclass[aps,pra,reprint,superscriptaddress,floatfix]{revtex4-2}

\usepackage{amsmath,amssymb,graphicx,hyperref,physics,orcidlink}
\usepackage[T1]{fontenc}
\usepackage[utf8]{inputenc}
\usepackage{lmodern}
\usepackage[dvipsnames]{xcolor}
\usepackage{booktabs,bm}

\hypersetup{
 colorlinks=true,
 linkcolor=blue,
 citecolor=blue,
 filecolor=blue,
 urlcolor=blue,
 breaklinks=true
}

\newcommand{\ii}{\mathrm{i}}
\newcommand{\ee}{\mathrm{e}}
\newcommand{\dif}{\mathrm{d}}
\newcommand{\kh}{\kappa_{\rm H}}
\newcommand{\xh}{x_{\rm H}}
\newcommand{\xs}{X_{\rm s}}
\newcommand{\ueff}{u_{\rm eff}}
\newcommand{\ve}{v_{\rm edge}}

\begin{document}

\date{\today}

\title{Dark-Soliton Branch Blocking in Transonic Bose--Einstein Condensate Flows}

\author{Edilberto O. Silva\,\orcidlink{0000-0002-0297-5747}}
\email{edilberto.silva@ufma.br}

\affiliation{Programa de P\'{o}s-Gradua\c{c}\~{a}o em F\'{i}sica, Universidade Federal do Maranh\~{a}o, 65080-805, S\~{a}o Lu\'{i}s, Maranh\~{a}o, Brazil}

\affiliation{Coordena\c{c}\~ao do Curso de F\'{\i}sica -- Bacharelado, Universidade Federal do Maranh\~{a}o, 65085-580 S\~{a}o Lu\'{\i}s, Maranh\~{a}o, Brazil}

\begin{abstract}
Acoustic horizons in Bose--Einstein condensates are usually characterized through long-wavelength Bogoliubov phonons. We study a nonlinear counterpart: whether a one-dimensional dark-soliton branch can sustain upstream laboratory motion in a stationary transonic flow. The mechanism is local at leading order. A regular dark soliton has a bounded fluid-frame velocity, limited by the local sound speed; therefore, in the supersonic region, the background flow exceeds the largest upstream velocity available to the soliton branch. The sonic point is thus the upstream edge of the local dark-soliton branch, rather than a hard wall or a soliton geodesic surface. We construct stationary transonic Gross--Pitaevskii backgrounds and evolve the full order parameter in an open, nonperiodic domain. The simulations show upstream propagation on the subsonic side, finite-depth stalling on the subsonic side, and downstream advection for defects initialized in the supersonic region with upstream velocity relative to the fluid. Convergence, branch-consistency, local-density, phase-jump, and a dense scan of dimensionless upstream attempts support the soliton-like interpretation. The result is a branch-existence constraint, not a rigorous lower bound on arbitrary density minima of the Gross--Pitaevskii field.
\end{abstract}

\maketitle

\section{Introduction}

Analog gravity in moving quantum fluids provides a controlled setting in which elements of curved-spacetime causal structure can be emulated in the laboratory \cite{Unruh1981,Barcelo2011}. In dilute Bose--Einstein condensates (BECs), the analogy arises by linearizing the Gross--Pitaevskii equation around a stationary flowing background. Long-wavelength Bogoliubov phonons then propagate in an effective acoustic geometry, and a horizon is formed where the background flow reaches the local sound speed \cite{Garay2000,Recati2009,Leonhardt2003}. This framework has motivated extensive studies of acoustic black holes, Hawking-like emission, mode conversion, and horizon instabilities in quantum fluids \cite{Steinhauer2016,Steinhauer2019}.

In the present context, the phrase ``acoustic medium'' has a precise meaning. It does not denote a separate classical sound-supporting material placed on top of the condensate. Rather, the acoustic medium is the hydrodynamic sector of the condensate itself: density and phase perturbations of the order parameter give rise to Bogoliubov sound waves propagating in a moving, compressible quantum fluid. The acoustic horizon is therefore a property of the stationary condensate background, located where the flow speed equals the local Bogoliubov sound speed. This point is useful because it allows one to ask whether the same transonic condensate configuration that blocks linear phonons also constrains finite-amplitude coherent matter-wave excitations.

The standard acoustic horizon is a linear concept. It is defined by the inability of low-energy phonons to propagate upstream once the flow becomes supersonic. The central question of this work is narrower and nonlinear: can a regular one-dimensional dark-soliton branch sustain upstream laboratory motion in a region where the background itself is supersonic? Dark solitons provide a natural probe of this question. They are localized density depletions carrying a phase jump; they have been generated and tracked experimentally in quasi-one-dimensional condensates \cite{Burger1999,Denschlag2000,Becker2008}, and they are among the best understood nonlinear excitations of repulsive Bose gases \cite{Kivshar1998,Frantzeskakis2010,Kevrekidis2015}. Unlike a phonon wave packet, a dark soliton has a coherent nonlinear core. Unlike a generic finite-amplitude disturbance, it belongs to a well-defined solution branch with a sharp endpoint: in the rest frame of a homogeneous condensate, the magnitude of the soliton velocity is bounded by the sound speed. The black soliton is stationary in the local fluid frame, while the shallow limit approaches the sound speed and merges with the phonon continuum.

Dark-soliton dynamics in slowly varying one-dimensional condensates has a long history. In particular, multiple-scale analyses show how grey solitons move in backgrounds whose density and velocity vary slowly on the scale of the soliton core \cite{BuschAnglin2000}. Dark solitons have also appeared in analogue-gravity contexts, for example as heavy nonlinear probes of Hawking-like fluctuations \cite{HangGabadadzeHuang2019}, and recent two-dimensional transonic-flow experiments observe self-oscillating supersonic dynamics associated with soliton emission inside acoustic horizons \cite{Tamura2025}. The present work is not meant to be the first connection between solitons and acoustic horizons. Its intended contribution is more specific: to formulate the elementary branch-blocking criterion for a one-dimensional dark soliton embedded in a stationary transonic Gross--Pitaevskii background, and to test that criterion by direct field evolution.

The mechanism is simple. Consider a stationary flow directed toward increasing $x$. A dark soliton trying to move upstream must have a negative velocity relative to the local fluid. Its laboratory velocity is the sum of the background velocity and this relative velocity, up to corrections from background gradients and radiation. Since a regular dark soliton cannot provide an arbitrarily large negative relative speed, the local soliton branch has no upstream laboratory-velocity member once the background flow is sonic or supersonic. The acoustic horizon is then inherited by the nonlinear sector not as a rigid wall, but as a branch edge of the dark-soliton family.

This claim is deliberately weaker than a full effective-particle theory. A dark soliton is not a point particle following null geodesics of the acoustic metric. A finite-depth soliton can turn before the horizon, deform in the near-horizon region, emit Bogoliubov radiation, or eventually lose a clean solitonic identity. The robust leading-order statement tested here is instead that, as long as the excitation remains a regular dark-soliton-like defect, it cannot maintain a persistent upstream laboratory branch in a region where the background flow exceeds the local sound speed. The quantitative trajectory, the emitted radiation, and the evolution of the depth must be determined by solving the full time-dependent Gross--Pitaevskii equation.

The practical motivation is twofold. First, a dark soliton is a localized phase-coherent probe. Its trajectory can distinguish qualitatively different regions of a transonic flow: subsonic upstream propagation, finite-depth stalling on the subsonic side, and unavoidable downstream advection on the supersonic side. This makes it complementary to phonon-based probes of acoustic horizons. Second, because solitons are matter-wave defects that can be generated, moved, and detected in guided condensates, the mechanism may be relevant for atomtronic or guided-matter-wave geometries in which one seeks to control nonlinear excitations \cite{Seaman2007,Pepino2009,Amico2021}. In such a setting, a transonic condensate can act as a directional filter for coherent nonlinear defects: the soliton branch itself supplies the speed limit.

The objective of this paper is therefore not to claim a new metric for solitons, but to isolate a nonlinear consequence of the same hydrodynamic sound cone that defines the acoustic horizon. Section~\ref{sec:model} presents the one-dimensional Gross--Pitaevskii model, its hydrodynamic form, and the construction of stationary transonic backgrounds. Section~\ref{sec:linear} recalls the linear acoustic horizon. Section~\ref{sec:soliton} reviews the homogeneous dark-soliton solution and its velocity bound. Section~\ref{sec:lda} states the local-density branch picture in a stationary flow and identifies the relevant adiabaticity parameter. Section~\ref{sec:separatrix} formulates the branch-blocking condition and clarifies the role of finite depth. Section~\ref{sec:energy} gives a deliberately heuristic energy--momentum perspective. Section~\ref{sec:numerics} describes the open-boundary numerical implementation and the soliton-identity diagnostics. Section~\ref{sec:results} presents the numerical evidence for the three regimes, branch-identity diagnostics, a dense dimensionless upstream-attempt scan, and convergence tests. Section~\ref{sec:discussion} discusses the physical interpretation, limitations, and possible experimental relevance.

\section{Gross--Pitaevskii model and stationary transonic backgrounds}
\label{sec:model}

We consider a quasi-one-dimensional repulsive condensate governed, in dimensionless units with $\hbar=m=1$, by the defocusing Gross--Pitaevskii equation
\begin{equation}
\ii\partial_t\psi(x,t)=
\left[-\frac{1}{2}\partial_x^2+V(x)+g(x)|\psi(x,t)|^2\right]\psi(x,t),
\label{eq:gpe}
\end{equation}
where $g(x)>0$ is the effective one-dimensional nonlinear coupling and $V(x)$ is an external potential. The order parameter is written in Madelung form as
\begin{equation}
\psi(x,t)=\sqrt{\rho(x,t)}\exp[\ii\theta(x,t)],
\label{eq:madelung}
\end{equation}
with density $\rho$ and velocity
\begin{equation}
v(x,t)=\partial_x\theta(x,t).
\label{eq:velocity}
\end{equation}
Substitution into Eq.~\eqref{eq:gpe} gives
\begin{equation}
\partial_t\rho+\partial_x(\rho v)=0,
\label{eq:continuity}
\end{equation}
and
\begin{equation}
\partial_t\theta+\frac{v^2}{2}+V(x)+g(x)\rho
-\frac{1}{2}\frac{\partial_x^2\sqrt{\rho}}{\sqrt{\rho}}=0.
\label{eq:bernoulli_time}
\end{equation}
The last term is the quantum-pressure contribution.

A stationary flowing background has the form
\begin{equation}
\psi_0(x,t)=\sqrt{\rho_0(x)}\exp[\ii\theta_0(x)-\ii\mu t],
\label{eq:background_order}
\end{equation}
where
\begin{equation}
v_0(x)=\partial_x\theta_0(x),
\qquad
J=\rho_0(x)v_0(x)
\label{eq:current}
\end{equation}
is a constant current. The stationary Bernoulli relation becomes
\begin{equation}
\mu=\frac{J^2}{2\rho_0^2}+V(x)+g(x)\rho_0
-\frac{1}{2}\frac{\partial_x^2\sqrt{\rho_0}}{\sqrt{\rho_0}}.
\label{eq:stationary_bernoulli}
\end{equation}
This equation may either be solved for $\rho_0(x)$ given $V(x)$ and $g(x)$, or inverted to design a stationary background. In the latter approach one prescribes a smooth density $\rho_0(x)$, coupling $g(x)$, current $J$, and chemical potential $\mu$, and defines
\begin{equation}
V(x)=\mu-\frac{J^2}{2\rho_0^2}-g(x)\rho_0
+\frac{1}{2}\frac{\partial_x^2\sqrt{\rho_0}}{\sqrt{\rho_0}}.
\label{eq:potential_from_background}
\end{equation}
The pair $(\rho_0,V)$ obtained in this way is not an arbitrary model landscape: it is constructed so that Eq.~\eqref{eq:background_order} is a stationary solution of Eq.~\eqref{eq:gpe}, up to numerical boundary and discretization errors.

The local speed of sound is
\begin{equation}
c_0(x)=\sqrt{g(x)\rho_0(x)}.
\label{eq:sound_speed}
\end{equation}
For a right-moving background flow, the acoustic horizon is located at
\begin{equation}
v_0(\xh)=c_0(\xh),
\label{eq:horizon_condition}
\end{equation}
with $v_0<c_0$ on the upstream subsonic side and $v_0>c_0$ on the downstream supersonic side. The associated horizon gradient is
\begin{equation}
\kh=\left.\partial_x\big[c_0(x)-v_0(x)\big]\right|_{x=\xh},
\label{eq:kappa}
\end{equation}
up to a sign convention. In the hydrodynamic long-wavelength limit, this quantity is proportional to the acoustic surface gravity.

A convenient family of backgrounds for numerical work is
\begin{equation}
\rho_0(x)=\rho_R+\frac{\rho_L-\rho_R}{2}
\left[1-\tanh\left(\frac{x-x_c}{\sigma_H}\right)\right],
\label{eq:rho_profile}
\end{equation}
with $\rho_L>\rho_R$. For constant $g=g_0$, the Mach number is
\begin{equation}
M(x)=\frac{v_0(x)}{c_0(x)}=\frac{J}{\sqrt{g_0}\rho_0^{3/2}(x)}.
\label{eq:mach}
\end{equation}
Thus a sonic point exists if $J$ lies between $\sqrt{g_0}\rho_R^{3/2}$ and $\sqrt{g_0}\rho_L^{3/2}$. The exact horizon position is determined by $M(\xh)=1$.

\section{Linear acoustic horizon}
\label{sec:linear}

Linearizing the order parameter as $\psi=\psi_0+\delta\psi$ gives the Bogoliubov spectrum in a local-density approximation,
\begin{equation}
\big[\omega-v_0(x)k\big]^2=c_0^2(x)k^2+\frac{k^4}{4}.
\label{eq:bogoliubov_dispersion}
\end{equation}
At long wavelength,
\begin{equation}
\omega\simeq v_0(x)k\pm c_0(x)k.
\label{eq:phonon_dispersion}
\end{equation}
The upstream branch has a laboratory group velocity
\begin{equation}
v_g^{\rm up}(x)\simeq v_0(x)-c_0(x),
\label{eq:phonon_group}
\end{equation}
for a right-moving flow. Equation~\eqref{eq:horizon_condition} is therefore the condition at which long-wavelength upstream phonons are blocked.

Equivalently, neglecting quantum pressure and working in $1+1$ dimensions, phase perturbations satisfy a wave equation in an effective acoustic metric whose line element may be written, up to a conformal factor, as
\begin{equation}
\dif s^2=-\big[c_0^2(x)-v_0^2(x)\big]\dif t^2
-2v_0(x)\dif t\dif x+\dif x^2.
\label{eq:acoustic_metric}
\end{equation}
The metric component multiplying $\dif t^2$ changes sign at $v_0=c_0$, identifying the acoustic horizon. The dark-soliton problem below is not obtained by treating the soliton as a phonon ray in this metric. Rather, the same combination $v_0-c_0$ emerges from the existence condition of the nonlinear defect.

\section{Homogeneous dark soliton and intrinsic speed bound}
\label{sec:soliton}

In a homogeneous condensate with density $\rho_b$, coupling $g_b$, sound speed
\begin{equation}
c_b=\sqrt{g_b\rho_b},
\label{eq:cb}
\end{equation}
and healing length
\begin{equation}
\xi_b=\frac{1}{\sqrt{g_b\rho_b}},
\label{eq:healing_length}
\end{equation}
the defocusing Gross--Pitaevskii equation admits the standard dark-soliton solution
\begin{align}
\psi_{\rm ds}(x,t)&=\sqrt{\rho_b}\,\ee^{-\ii\mu t}
\notag\\
&\times
\left[
\ii\frac{u}{c_b}
+\sqrt{1-\frac{u^2}{c_b^2}}
\tanh\left(\frac{x-X(t)}{\xi_b/\sqrt{1-u^2/c_b^2}}\right)
\right],
\label{eq:dark_soliton}
\end{align}
where $u=\dot X$ is the soliton velocity in the fluid rest frame. A regular dark soliton exists only for
\begin{equation}
|u|<c_b.
\label{eq:local_bound_homogeneous}
\end{equation}
The black soliton corresponds to $u=0$. As $|u|\to c_b$, the minimum density approaches the background density, the soliton broadens, and the defect becomes indistinguishable from the sound-wave continuum.

The depth and phase jump are controlled by the same ratio $u/c_b$. The minimum density is
\begin{equation}
\rho_{\min}=\rho_b\frac{u^2}{c_b^2},
\label{eq:minimum_density}
\end{equation}
and the density contrast is
\begin{equation}
D=1-\frac{\rho_{\min}}{\rho_b}=1-\frac{u^2}{c_b^2}.
\label{eq:depth_speed_relation}
\end{equation}
The phase jump across the soliton is
\begin{equation}
\Delta\theta=2\arccos\left(\frac{u}{c_b}\right),
\label{eq:phase_jump}
\end{equation}
up to the convention used for the propagation direction. The velocity bound is therefore not an independent assumption; it is built into the soliton branch itself.

\section{Local-density soliton branch in a stationary flow}
\label{sec:lda}

For a slowly varying background, the soliton may be treated as a localized defect whose parameters adjust to the local density and sound speed. A useful ansatz is
\begin{align}
\psi(x,t)&\approx \sqrt{\rho_0(x)}
\ee^{\ii\theta_0(x)-\ii\mu t+\ii\chi(x,t)}
\notag\\
&\quad\times
\left[
\ii\sin\phi(t)+\cos\phi(t)
\tanh\left(\frac{x-X(t)}{\ell(t)}\right)
\right].
\label{eq:local_ansatz}
\end{align}
Here $X(t)$ is the soliton center, $\phi(t)$ is a grayness parameter, $\ell(t)$ is the width, and $\chi(x,t)$ accounts for smooth backflow and radiative corrections. To the leading local-density order,
\begin{equation}
u(X,\phi)=c_0(X)\sin\phi,
\label{eq:u_phi}
\end{equation}
\begin{equation}
\ell(X,\phi)=\frac{\xi(X)}{\cos\phi},
\qquad
\xi(X)=\frac{1}{\sqrt{g(X)\rho_0(X)}}.
\label{eq:ell_phi}
\end{equation}
The local existence condition is again
\begin{equation}
|u(X,\phi)|<c_0(X).
\label{eq:local_bound}
\end{equation}
In the laboratory frame, the center velocity has the form
\begin{equation}
\dot X=v_0(X)+u(X,\phi)+\delta v_{\rm grad}+\delta v_{\rm rad}.
\label{eq:collective_velocity}
\end{equation}
The term $\delta v_{\rm grad}$ denotes corrections due to finite background gradients, while $\delta v_{\rm rad}$ accounts for momentum exchange with emitted Bogoliubov radiation. These corrections are quantitative and must be resolved dynamically near a sharp horizon. They also imply that the branch criterion below should not be advertised as a rigorous theorem for every density minimum of the full field. It is the leading local-density statement for a regular dark-soliton branch.

The ansatz is not used below as a numerical constraint. It serves only to make explicit the local soliton family whose fluid-frame velocity is bounded by $c_0(X)$. In the simulations, the field is evolved directly from Eq.~\eqref{eq:gpe_rotating}; the trajectory and depth are then extracted from the resulting density. Thus the local-density argument provides the kinematic expectation, while the Gross--Pitaevskii evolution tests whether the soliton-like object actually follows it in the presence of gradients and radiation.

The shortest hydrodynamic scale is often characterized by
\begin{equation}
\epsilon_{\xi}=\frac{\xi(X)}{L_{\rm bg}(X)},
\label{eq:lda_parameter_xi}
\end{equation}
with
\begin{equation}
L_{\rm bg}^{-1}(X)=\max\left\{\left|\partial_x\ln\rho_0\right|,
\left|\partial_x\ln c_0\right|,
\left|\partial_x\ln v_0\right|\right\}_{x=X}.
\label{eq:lda_length}
\end{equation}
For a grey soliton, however, the relevant object is not just the healing length but the soliton width. Using Eq.~\eqref{eq:depth_speed_relation},
\begin{equation}
\ell(X)=\frac{\xi(X)}{\sqrt{D(X)}}
=\frac{\xi(X)}{\sqrt{1-u^2(X)/c_0^2(X)}}.
\label{eq:ell_depth}
\end{equation}
The more stringent local-density control parameter is therefore
\begin{equation}
\epsilon_{\rm sol}(X)=\frac{\ell(X)}{L_{\rm bg}(X)}
=\frac{\xi(X)}{L_{\rm bg}(X)\sqrt{D(X)}}.
\label{eq:epsilon_sol}
\end{equation}
This quantity diverges in the shallow limit $|u|\to c_0$, precisely where the soliton approaches the phonon continuum. Consequently, the branch-edge criterion should be interpreted with care near $D\to0$: the local existence interval remains the correct leading-order guide, but quantitative trajectories and loss of contrast require direct field evolution.

\section{Branch separatrix and finite-depth turning points}
\label{sec:separatrix}

Consider a right-moving background flow. An upstream-moving soliton has negative fluid-frame velocity. At leading local-density order, the regular soliton branch occupies the velocity interval
\begin{equation}
-c_0(X)<u(X)<c_0(X).
\label{eq:u_interval}
\end{equation}
Neglecting $\delta v_{\rm grad}$ and $\delta v_{\rm rad}$ only for the purpose of identifying the branch edge, Eq.~\eqref{eq:collective_velocity} gives
\begin{equation}
v_0(X)-c_0(X)<\dot X<v_0(X)+c_0(X).
\label{eq:lab_interval}
\end{equation}
The upstream edge of the local branch is therefore
\begin{equation}
\ve(X)=v_0(X)-c_0(X).
\label{eq:branch_edge}
\end{equation}
This is the central theoretical statement. It is not a hard dynamical lower bound on an arbitrary depression of the density. It is the lower edge of the adiabatic dark-soliton velocity branch.

The branch-edge criterion gives
\begin{widetext}
\begin{align}
&v_0(X)<c_0(X) &&\Rightarrow \quad \text{a local upstream dark-soliton branch can exist},
\label{eq:subsonic_possible}\\
&v_0(X)=c_0(X) &&\Rightarrow \quad \text{the maximal upstream soliton speed is exhausted},
\label{eq:horizon_saturation}\\
&v_0(X)>c_0(X) &&\Rightarrow \quad \text{no regular local dark-soliton branch has upstream laboratory velocity}.
\label{eq:supersonic_forbidden}
\end{align}
\end{widetext}
Thus the acoustic horizon is a limiting separatrix of the dark-soliton branch because the soliton cannot provide enough upstream relative velocity to overcome a sonic or supersonic background flow. The statement specifies where a sustained upstream branch can exist, not the exact time at which a finite-depth soliton must turn, deform, radiate, or lose soliton identity.

For a finite-depth soliton, the actual turning point need not coincide with the acoustic horizon. A local turning point satisfies
\begin{equation}
v_0(X_*)+u(X_*,\phi_*)=0,
\label{eq:finite_depth_turning}
\end{equation}
or
\begin{equation}
v_0(X_*)=|u(X_*,\phi_*)|.
\label{eq:turning_abs}
\end{equation}
Because $|u|<c_0$, Eq.~\eqref{eq:turning_abs} can be satisfied on the subsonic side of the horizon. Deep solitons, which have small $|u|$, should turn or stall farther from the horizon. Shallow solitons, which have $|u|$ closer to $c_0$, can approach the sonic point more closely, but their width also grows according to Eq.~\eqref{eq:ell_depth}. The horizon is therefore the limiting turning surface of the soliton branch, not necessarily the reflection point of every individual soliton.

This distinction is crucial for the interpretation of simulations. Reflection before the horizon, finite-depth stalling on the subsonic side, deformation near the horizon, downstream advection, and loss of soliton identity are all compatible with the same branch constraint. What is excluded at leading local-density order is a regular dark soliton maintaining a persistent upstream laboratory branch in a region where $v_0>c_0$.

\section{Heuristic energy--momentum perspective}
\label{sec:energy}

The branch criterion above does not require an energy argument. Nevertheless, the homogeneous dark-soliton energy--momentum relation gives useful intuition. In a homogeneous condensate, the energy of a dark soliton relative to the background is
\begin{equation}
E_{\rm ds}=\frac{4}{3}\rho_b c_b
\left(1-\frac{u^2}{c_b^2}\right)^{3/2},
\label{eq:soliton_energy}
\end{equation}
while one standard expression for the canonical momentum is
\begin{equation}
P_{\rm ds}=2\rho_b\left[
\arccos\left(\frac{u}{c_b}\right)-
\frac{u}{c_b}\sqrt{1-\frac{u^2}{c_b^2}}
\right],
\label{eq:soliton_momentum}
\end{equation}
up to the usual branch choice and background contribution. A moving inhomogeneous condensate may be described locally by a Doppler-shifted collective Hamiltonian of the schematic form
\begin{align}
E_{\rm lab}(X,u)&\simeq
E_{\rm ds}[\rho_0(X),c_0(X),u]
\notag\\
&\quad
+v_0(X)P_{\rm ds}[\rho_0(X),c_0(X),u]
\notag\\
&\quad
+\text{background terms}.
\label{eq:lab_energy}
\end{align}
Because a dark soliton carries a particle deficit and because $P_{\rm ds}$ is convention- and branch-dependent, Eq.~\eqref{eq:lab_energy} is not used here as a derivation of the blocking condition. Its role is only heuristic. The regular soliton branch terminates at $|u|=c_0(X)$, where the contrast vanishes and the width diverges. If upstream laboratory motion in a supersonic right-moving flow would require $u<-v_0(X)<-c_0(X)$, no member of the regular local branch can supply that relative velocity.

The same viewpoint clarifies why near-branch-edge dynamics can be nonadiabatic. As a soliton approaches the shallow limit, its contrast decreases and its width grows. At the same time, inhomogeneity and background acceleration can couple the defect to Bogoliubov modes. A fully dynamical treatment must therefore track both the coherent soliton core and the emitted radiation.

\section{Numerical implementation and diagnostics}
\label{sec:numerics}

The branch argument above is deliberately local. To test whether it survives in the full nonlinear dynamics, we solve the time-dependent Gross--Pitaevskii equation for the complex field itself and extract the soliton trajectory, depth, and velocity from the evolved density. No trajectory, depth, width, residual field, or detector signal is imposed by hand.

Because the stationary transonic background is nonperiodic, with different asymptotic densities on the two sides, the production runs are performed in an open domain rather than in a periodic Fourier box. We evolve the equation in the rotating frame,
\begin{equation}
\ii\partial_t\psi=
\left[-\frac{1}{2}\partial_x^2+V(x)+g|\psi|^2-\mu\right]\psi,
\label{eq:gpe_rotating}
\end{equation}
using centered finite differences for $\partial_x^2$ and a fourth-order Runge--Kutta time step. The stationary background in this frame is
\begin{equation}
\psi_{\rm bg}(x)=\sqrt{\rho_0(x)}\ee^{\ii\theta_0(x)}.
\label{eq:psi_bg_rotating}
\end{equation}
The two boundary values are pinned to $\psi_{\rm bg}$, and a sponge layer damps only deviations from the stationary background,
\begin{align}
\partial_t\psi
&=
-\ii\left[-\frac{1}{2}\partial_x^2+V(x)+g|\psi|^2-\mu\right]\psi
\notag\\
&\quad
-\Gamma(x)\left[\psi-\psi_{\rm bg}(x)\right].
\label{eq:sponge}
\end{align}
This form is essential. Damping the full field would deplete the current-carrying stationary state; damping only $\psi-\psi_{\rm bg}$ leaves the transonic background as a fixed point of the open-boundary evolution up to finite-difference error.

The background is generated from Eq.~\eqref{eq:rho_profile} with constant $g=g_0$ and constant current $J$. The potential $V(x)$ is computed from Eq.~\eqref{eq:potential_from_background}. In the runs reported below we use $L=520$, $N=6000$, $\Delta t=10^{-3}$, and $t_{\max}=90$. The Mach number crosses unity at
\begin{equation}
\xh\simeq 41.
\label{eq:xh_num}
\end{equation}
The stationary residual is quantified by
\begin{equation}
\epsilon_R=
\frac{
\left[\int_{\Omega_{\rm phys}}
\left|
\left(-\frac{1}{2}\partial_x^2+V+g|\psi_{\rm bg}|^2-\mu\right)
\psi_{\rm bg}
\right|^2\dif x\right]^{1/2}
}{
\left[\int_{\Omega_{\rm phys}}|\psi_{\rm bg}|^2\dif x\right]^{1/2}
},
\label{eq:stationary_residual}
\end{equation}
where $\Omega_{\rm phys}$ excludes the sponge layers. We obtain
\begin{equation}
\epsilon_R=1.61\times10^{-4},
\label{eq:residual_value}
\end{equation}
and the background-only evolution remains stable over the full simulation time, with density deviations of order $10^{-4}$.

The simulation scripts used to generate the figures specify the background constants $(\rho_L,\rho_R,J,g_0,x_c,\sigma_H,\mu)$, the physical interval, the sponge profile $\Gamma(x)$, and the tracking-window parameters. The equations above specify the construction, while the numerical values $L$, $N$, $\Delta t$, $t_{\max}$, $\xh$, and $\epsilon_R$ specify the production discretization and validation used in the present runs.

\begin{figure}[tbhp]
\centering
\includegraphics[scale=0.4]{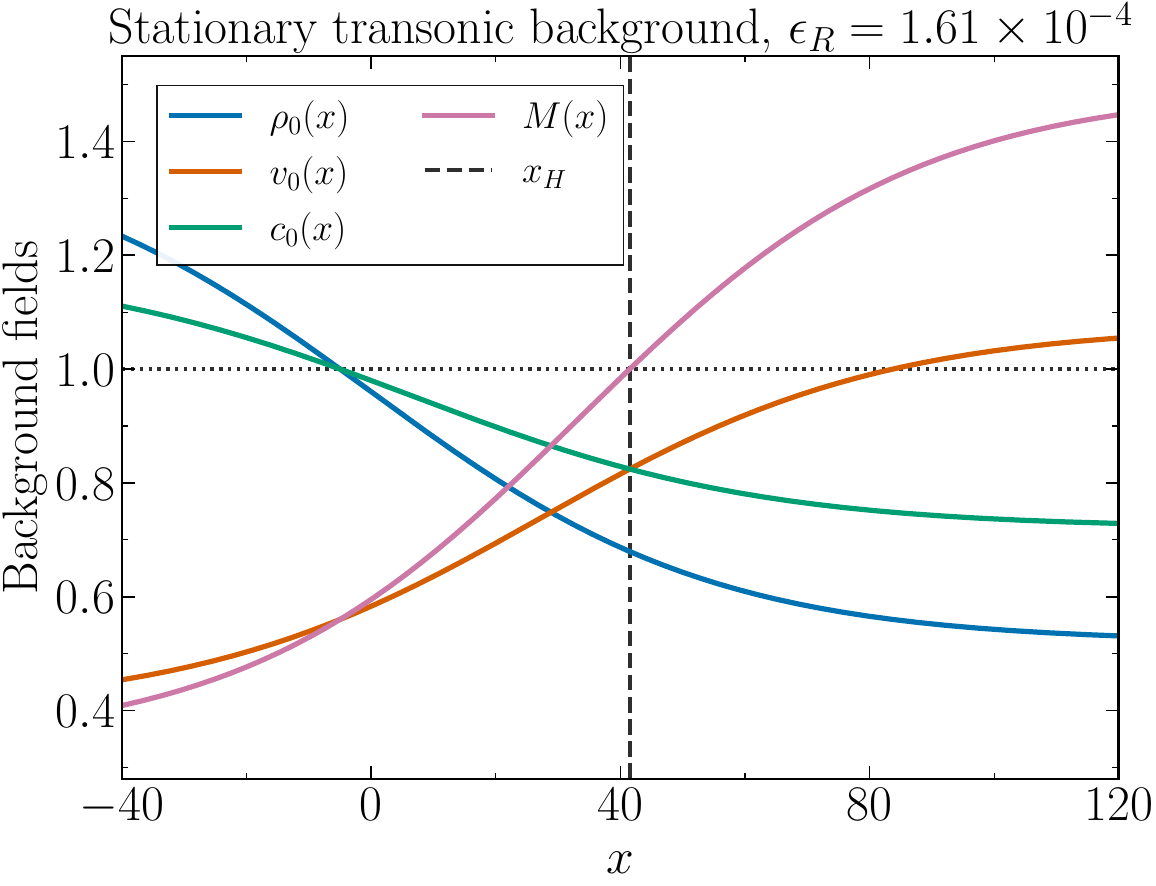}
\caption{Stationary transonic background used in the numerical simulations. The density $\rho_0(x)$ decreases from left to right, while the conservation law $v_0=J/\rho_0$ accelerates the flow. The local sound speed is $c_0=\sqrt{g\rho_0}$, and the Mach number $M=v_0/c_0$ crosses unity at the acoustic horizon $x_H$. The stationary residual is $\epsilon_R=1.61\times10^{-4}$.}
\label{fig:background}
\end{figure}

The initial dark soliton is obtained by multiplying the background field by a local homogeneous soliton profile. At the initial position $X_0$ we define
\begin{equation}
c_s=c_0(X_0),\qquad
\xi_s=\frac{1}{\sqrt{g(X_0)\rho_0(X_0)}} ,
\label{eq:local_initial_params_num}
\end{equation}
and use
\begin{align}
\psi(x,0)&=\sqrt{\rho_0(x)}\ee^{\ii\theta_0(x)}
\notag\\
&\quad\times
\left[
\ii\frac{u_0}{c_s}
+\sqrt{1-\frac{u_0^2}{c_s^2}}\,
\tanh\left(
\frac{x-X_0}{\xi_s/\sqrt{1-u_0^2/c_s^2}}
\right)
\right].
\label{eq:initial_soliton}
\end{align}
The flow is toward increasing $x$, and upstream attempts correspond to $u_0<0$. We perform three complementary scans. In the first, the soliton is placed on the subsonic side, $X_0=20$, with $u_0=-0.20,-0.40,-0.60,-0.75$. In the second, the soliton is placed in the supersonic region, $X_0=70$, with $u_0=-0.20,-0.40,-0.60$. These two fixed-$u_0$ scans establish the three representative regimes discussed below. In the third, we keep the launch point fixed in the supersonic region, $X_0=70$, and parameterize the initial upstream attempt by
\begin{equation}
\alpha=-\frac{u_0}{c_0(X_0)}.
\label{eq:alpha_scan_definition}
\end{equation}
We use $\alpha=0.50,0.60,0.70,0.80,0.90,0.95,0.98$, so that the last cases approach the local sound-speed endpoint of the dark-soliton branch. This targeted scan is not a full two-parameter phase diagram, but it directly tests whether the representative supersonic advection result survives as the initial fluid-frame velocity approaches the largest upstream value allowed by the local branch.

At each saved time, the soliton center is extracted as the density minimum inside a window tied to the previous position,
\begin{equation}
\xs(t)=\arg\min_{x\in{\cal W}(t)}|\psi(x,t)|^2.
\label{eq:center_extraction}
\end{equation}
This procedure is deliberately local: it follows the coherent density depletion rather than imposing an effective-particle equation. The trajectory is smoothed only in post-processing before computing $\dot{\xs}(t)$, since numerical differentiation amplifies small acoustic oscillations. The smoothing is therefore a diagnostic operation, not part of the dynamics. The branch edge is evaluated along the extracted trajectory as
\begin{equation}
\ve(t)=v_0[\xs(t)]-c_0[\xs(t)].
\label{eq:bound_along_traj}
\end{equation}
The soliton depth is measured from the local density depletion relative to a smooth background estimate,
\begin{equation}
D(t)=1-\frac{|\psi[\xs(t),t]|^2}{n_{\rm bg}[\xs(t),t]},
\label{eq:depth_diag}
\end{equation}
where $n_{\rm bg}$ is estimated from the density away from the core inside the same tracking window. Because the background is inhomogeneous and weak sound is emitted during the initial adjustment, $D(t)$ should be interpreted as an effective tracked depletion rather than as an exactly conserved homogeneous-soliton parameter.

A useful soliton-identity diagnostic is obtained from the effective fluid-frame velocity
\begin{equation}
\ueff(t)=\dot\xs(t)-v_0[\xs(t)].
\label{eq:ueff}
\end{equation}
For a perfectly local homogeneous soliton, the extracted velocity, depth, and phase jump would satisfy
\begin{equation}
D_{\rm branch}(t)=1-\frac{\ueff^2(t)}{c_0^2[\xs(t)]},
\label{eq:D_branch}
\end{equation}
\begin{equation}
\Delta\theta_{\rm branch}(t)=2\arccos\left(\frac{\ueff(t)}{c_0[\xs(t)]}\right),
\label{eq:theta_branch}
\end{equation}
when $|\ueff|<c_0$. Equations~\eqref{eq:D_branch} and \eqref{eq:theta_branch} define a nontrivial consistency diagnostic of soliton identity. Below we evaluate this first branch-consistency check for the three representative cases, together with the width-based local-density parameter $\epsilon_{\rm sol}(t)$. Deviations from the homogeneous relations are expected in an inhomogeneous open flow and are interpreted as nonadiabatic, radiative, and tracking corrections, not as an imposed soliton law.

\section{Numerical results: three dynamical regimes}
\label{sec:results}

\begin{figure}[tbhp]
\centering
\includegraphics[scale=0.4]{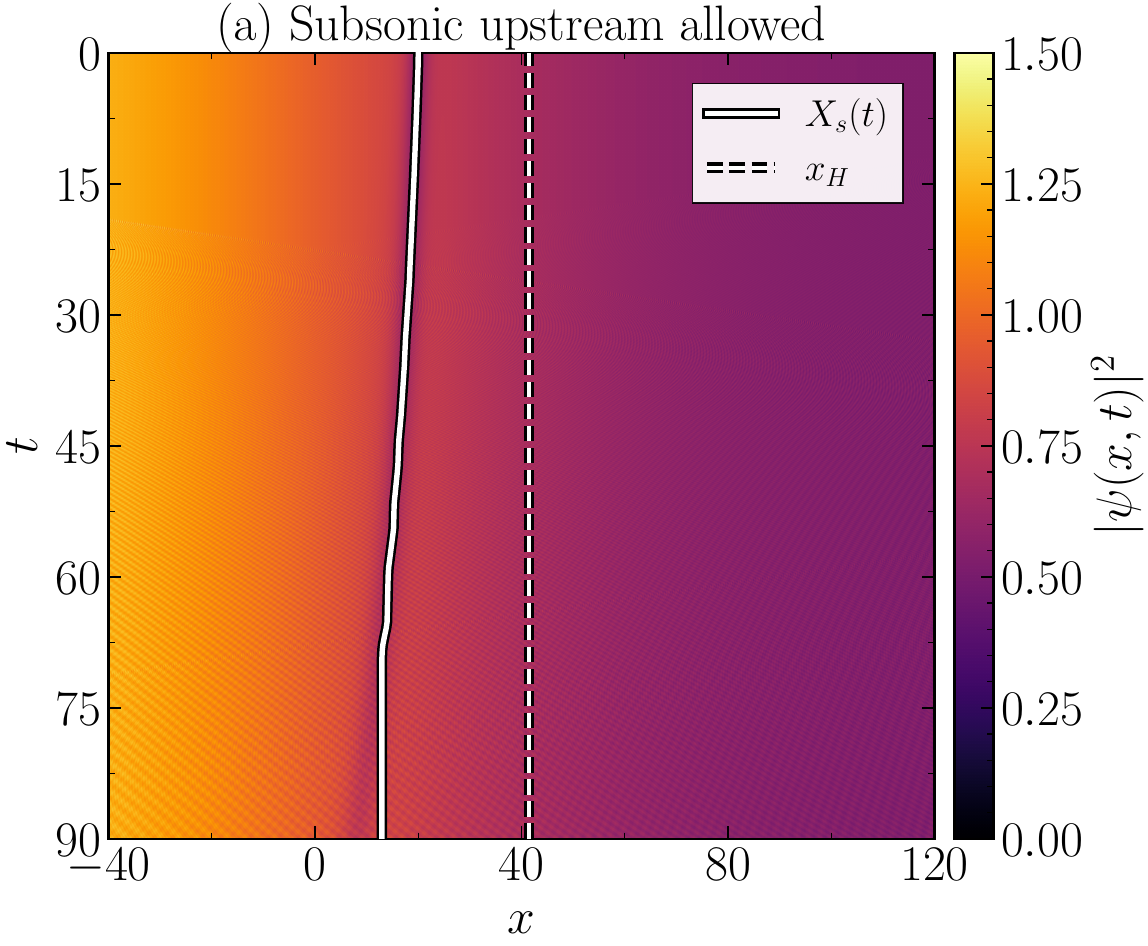}
\includegraphics[scale=0.4]{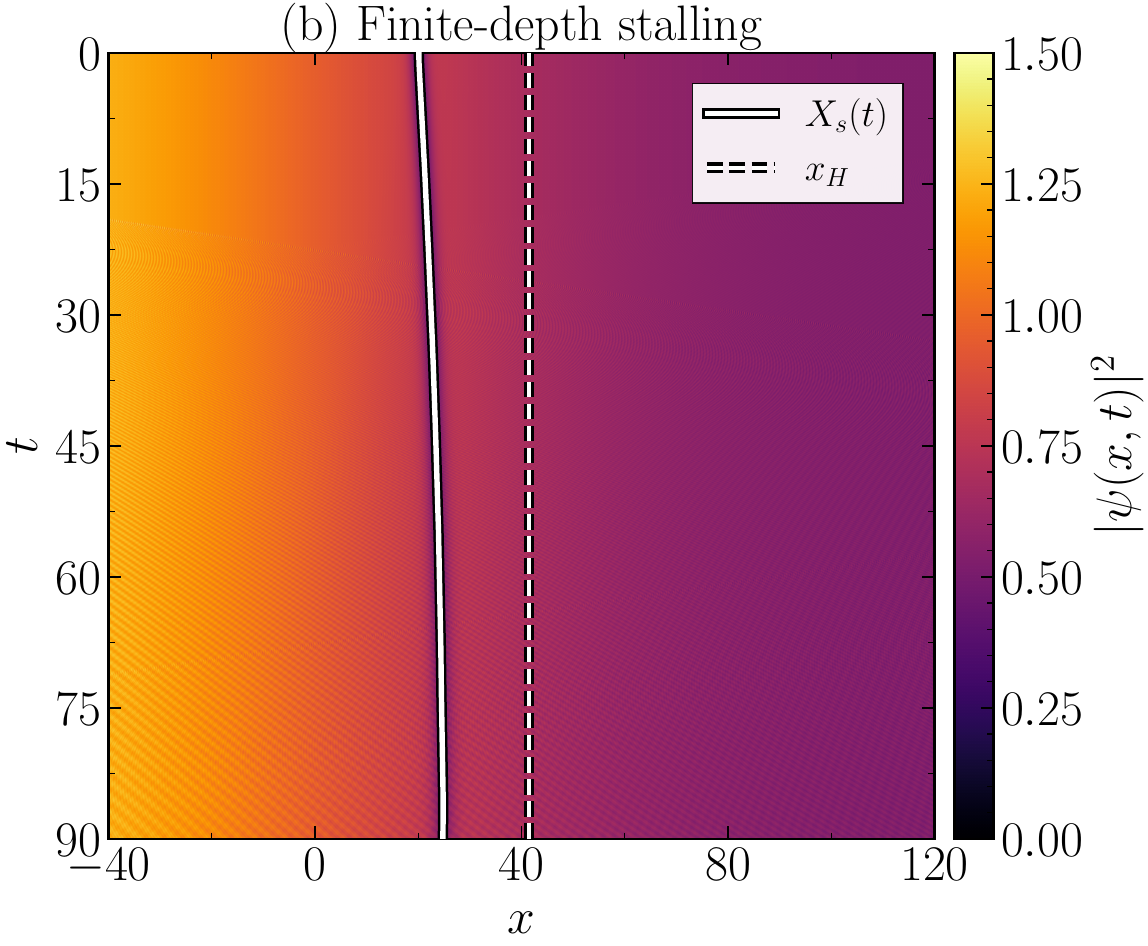}
\includegraphics[scale=0.4]{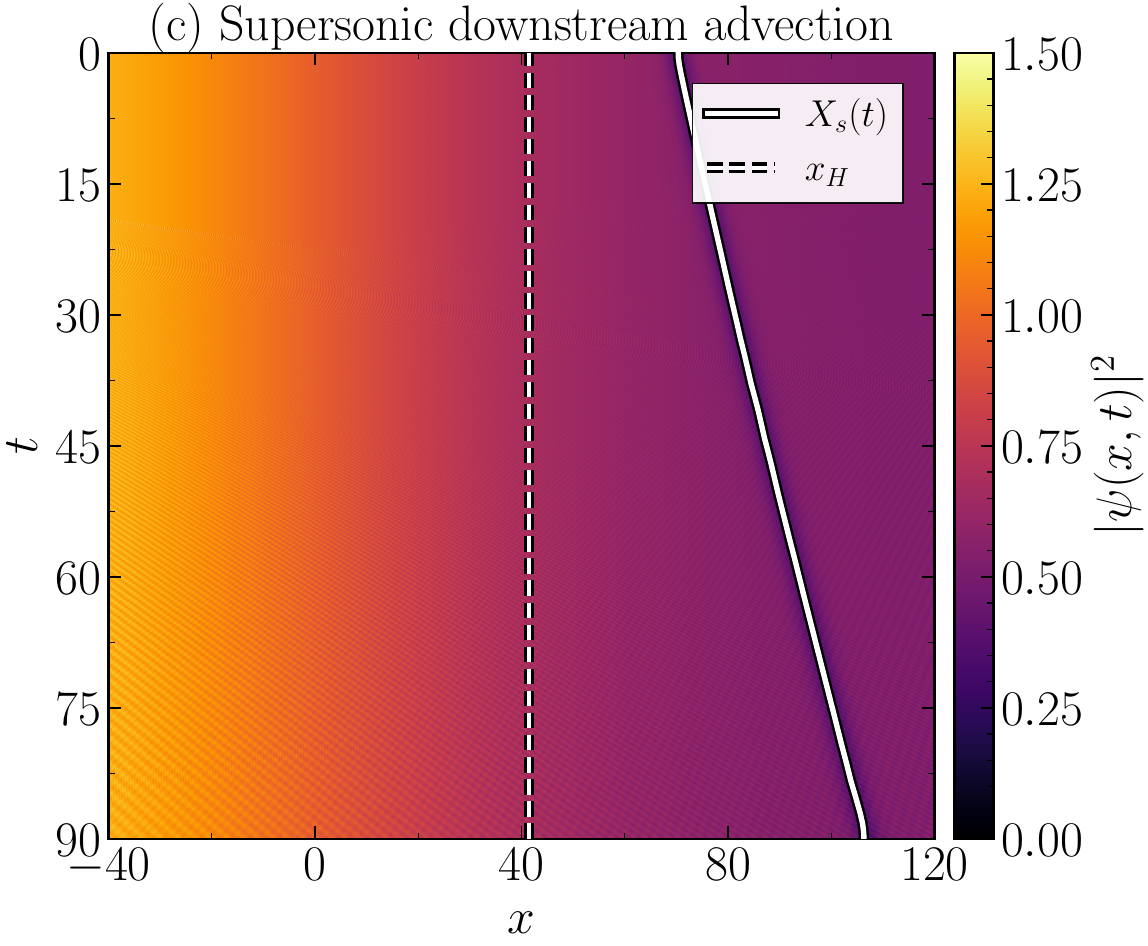}
\caption{Representative density evolutions $|\psi(x,t)|^2$ for the three dynamical regimes. The solid curve in each panel is the extracted trajectory $X_s(t)$, and the dashed vertical line marks the acoustic horizon $x_H$. (a) Subsonic upstream propagation for $X_0=20$ and $u_0=-0.75$: the soliton remains on the subsonic side and moves toward smaller $x$. (b) Finite-depth stalling on the subsonic side for $X_0=20$ and $u_0=-0.60$: the soliton is initially directed upstream in the local fluid frame but drifts slowly downstream because its relative velocity is insufficient to overcome the local convective flow. (c) Supersonic downstream advection for $X_0=70$ and $u_0=-0.60$: even though the initial velocity is upstream in the local fluid frame, the laboratory trajectory moves toward increasing $x$.}
\label{fig:density_regimes}
\end{figure}

\begin{figure}[tbhp]
\centering
\includegraphics[scale=0.4]{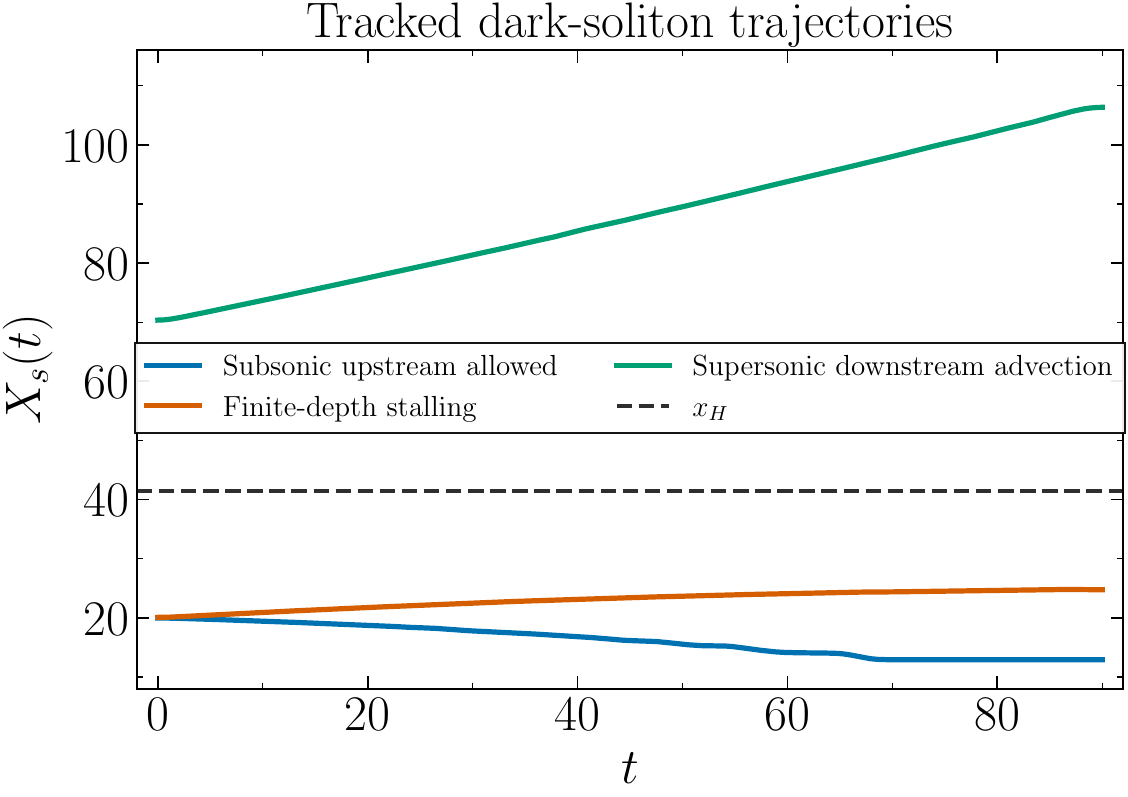}
\caption{Tracked soliton trajectories for the three representative regimes. In the subsonic upstream case, a sufficiently large negative relative velocity produces motion toward smaller $x$. A finite-depth subsonic attempt is nearly stalled and then drifts downstream. In the supersonic region the trajectory is strictly downstream, illustrating the absence of a sustained upstream dark-soliton branch when $v_0>c_0$.}
\label{fig:trajectories}
\end{figure}

\begin{figure}[tbhp]
\centering
\includegraphics[scale=0.4]{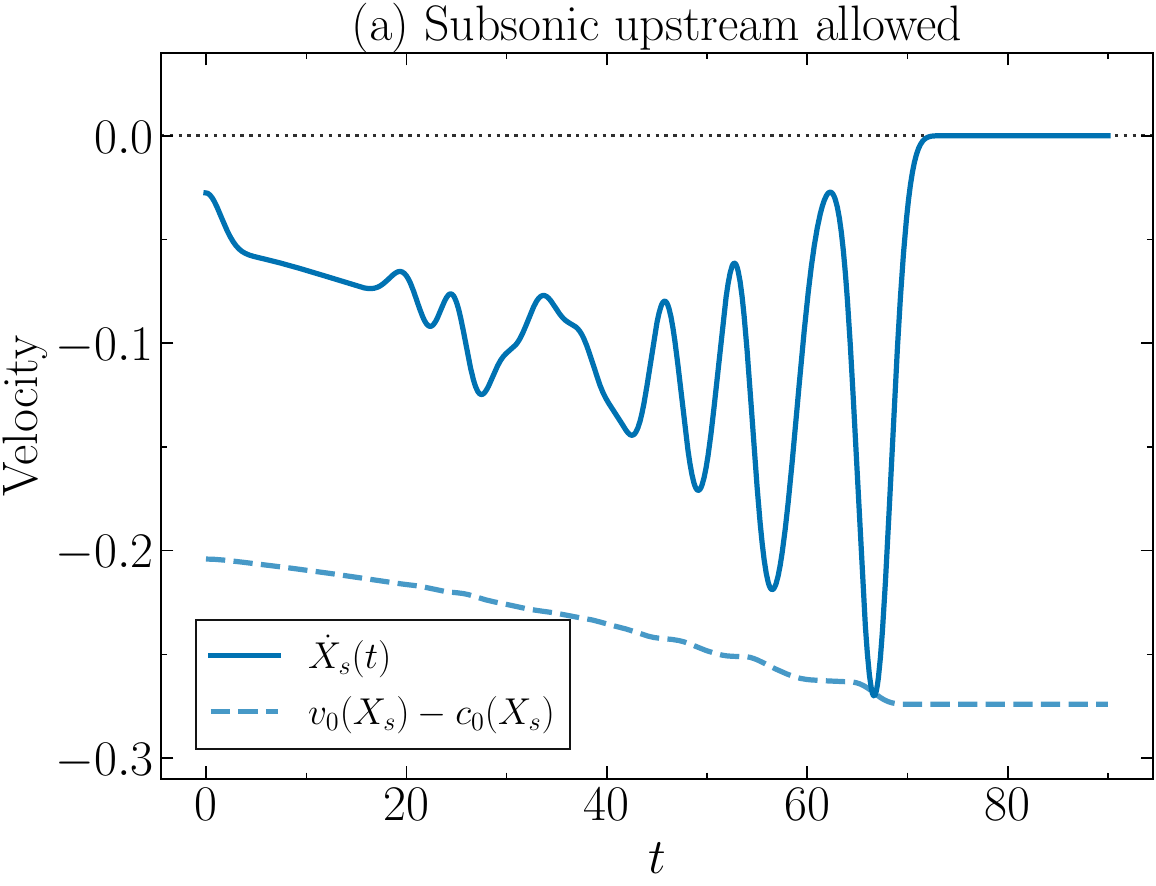}
\includegraphics[scale=0.4]{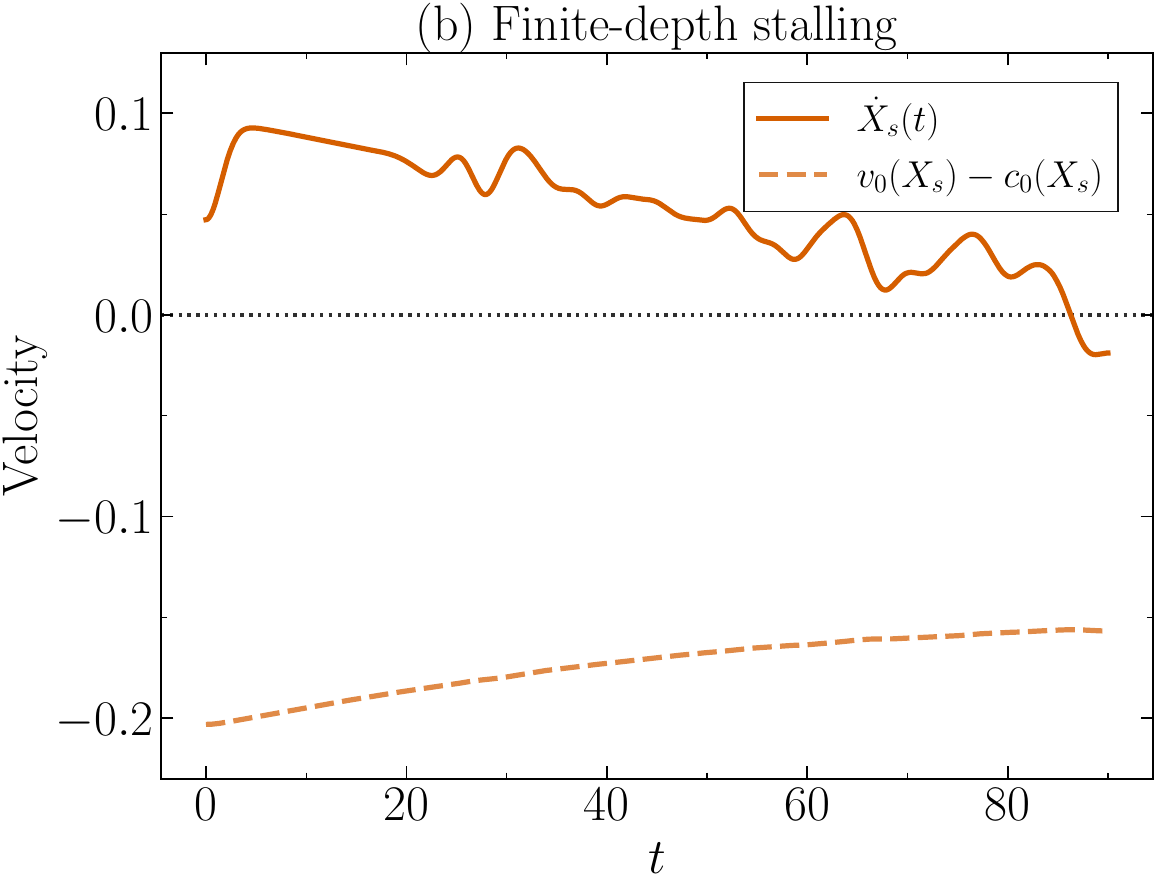}
\includegraphics[scale=0.4]{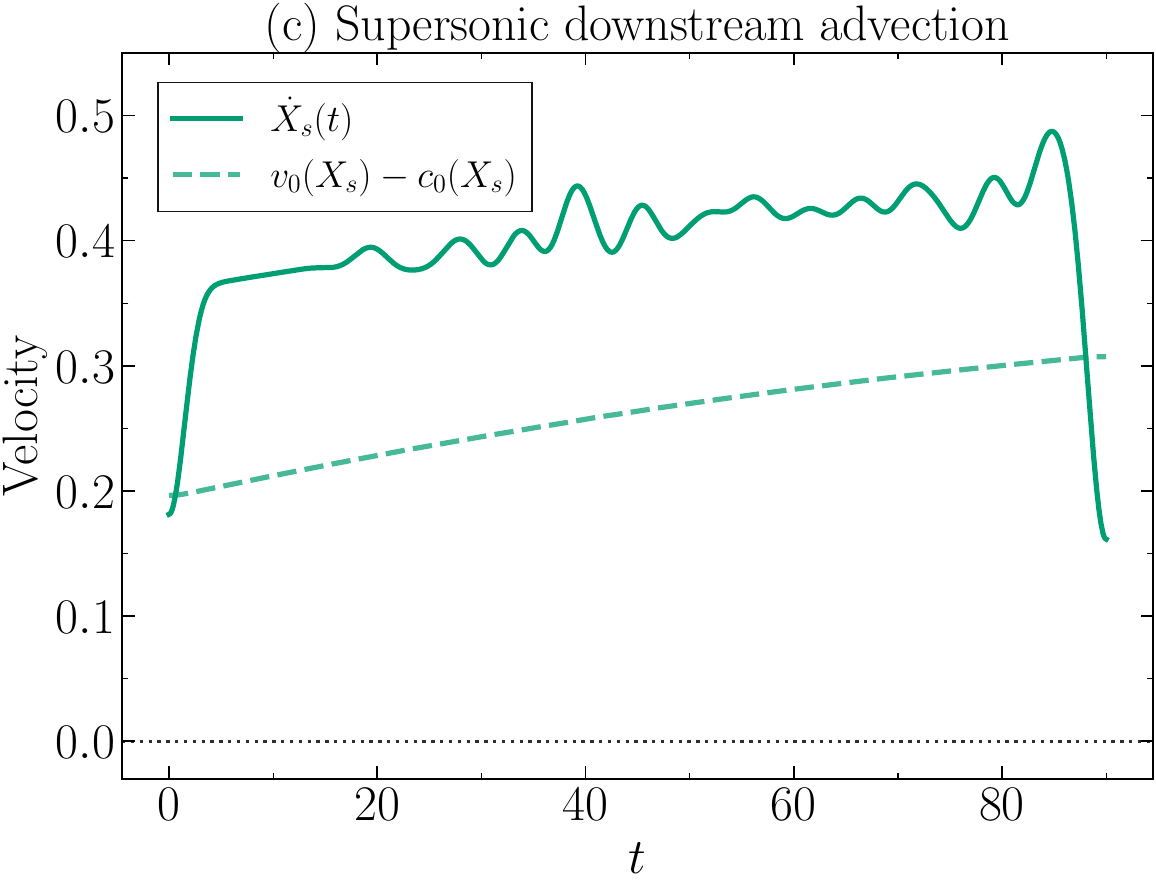}
\caption{Velocity-branch diagnostics for the three representative regimes. The solid curve in each panel is the smoothed soliton velocity $\dot X_s(t)$ extracted from the tracked density minimum, while the dashed curve is the local adiabatic branch edge $v_0[X_s(t)]-c_0[X_s(t)]$. (a) In the subsonic upstream-propagation case, the branch edge is negative, and the soliton can realize upstream laboratory motion. (b) In the finite-depth stalling case, the branch edge remains negative, but the finite-depth soliton drifts weakly downstream and stays close to the separatrix. (c) In the supersonic downstream-advection case, the branch edge is positive, and the extracted soliton velocity remains positive.}
\label{fig:velocity_bound}
\end{figure}

\begin{figure}[tbhp]
\centering
\includegraphics[scale=0.4]{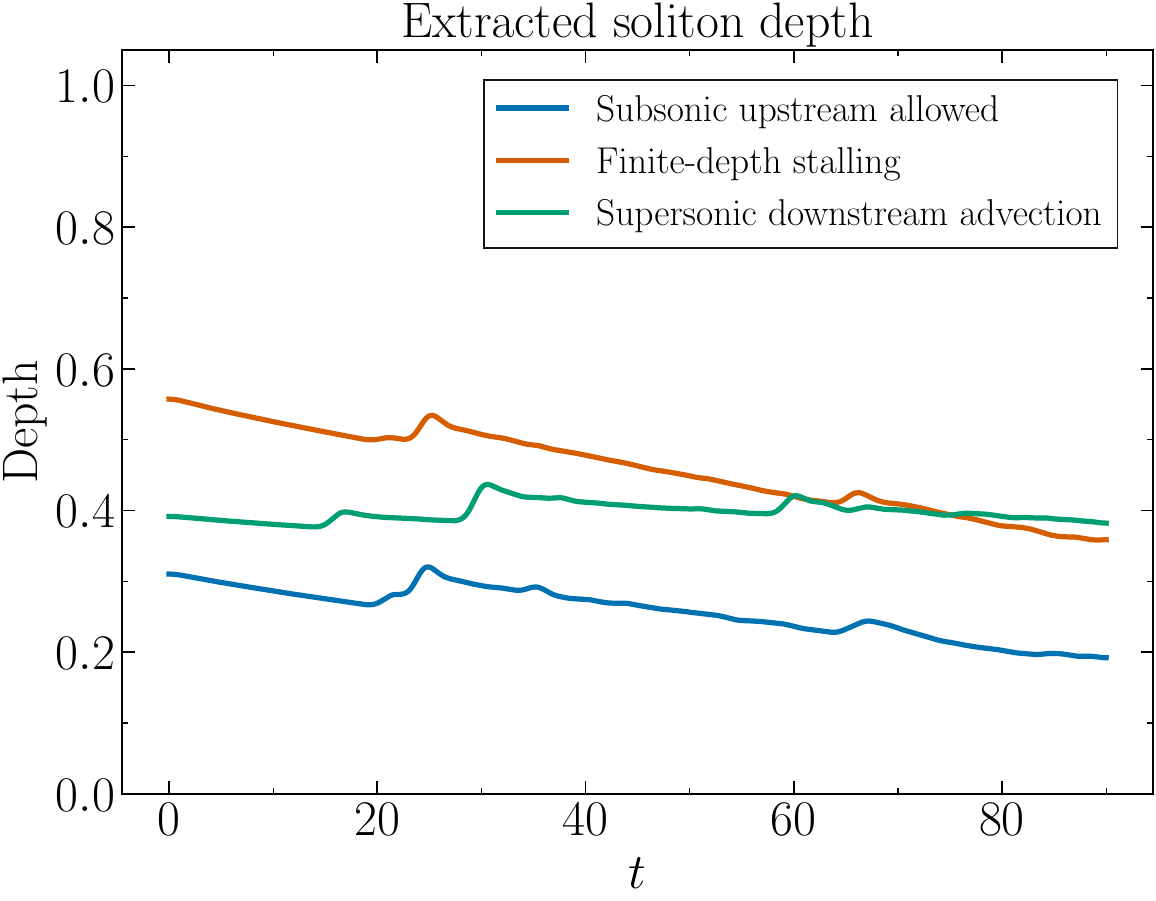}
\caption{Extracted soliton depth for the three representative regimes. The depth evolves slowly due to background inhomogeneity and weak radiation. The tracked density depletion remains a coherent, soliton-like object during the interval used to test the branch criterion while exhibiting the expected deformation in the inhomogeneous flow.}
\label{fig:depth}
\end{figure}

The simulations reveal three regimes, all obtained from the same stationary background and the same Gross--Pitaevskii evolution. The representative cases are summarized in Table~\ref{tab:regimes}. The corresponding density evolutions are shown in Fig.~\ref{fig:density_regimes}(a)--\ref{fig:density_regimes}(c), while the extracted trajectories, velocity-branch diagnostics, and depth evolution are shown in Figs.~\ref{fig:trajectories}, \ref{fig:velocity_bound}, and \ref{fig:depth}.

The three cases are not meant to define three separate models. They are three initial conditions in the same transonic condensate. Their role is to separate the physical mechanisms: a sufficiently fast upstream attempt in the subsonic region, a finite-depth attempt that stalls on the subsonic side, and an upstream attempt launched on the supersonic side. This organization makes the central prediction falsifiable: if a regular dark soliton could sustain upstream motion in the supersonic region, the branch-blocking interpretation would fail.

The three regimes should be viewed as nonlinear scattering outcomes of the same microscopic equation on the same transonic background. No force law for the soliton center is imposed; the trajectory is extracted a posteriori from the evolved density field.

\begin{table}[tbhp]
\caption{Representative upstream tests. The quoted velocities are mean values extracted from the smoothed trajectory over the simulation interval. The second row is a finite-depth subsonic stalling case; it should not be interpreted as a direct collision with the horizon.}
\label{tab:regimes}
\begin{ruledtabular}
\begin{tabular}{lcccc}
Regime & $X_0$ & $u_0$ & $X_s(t_{\max})$ & $\langle \dot X_s\rangle$\\
\hline
Subsonic upstream allowed & $20$ & $-0.75$ & $12.94$ & $-0.080$\\
Subsonic finite-depth stalling & $20$ & $-0.60$ & $24.76$ & $0.053$\\
Supersonic advection & $70$ & $-0.60$ & $106.33$ & $0.404$
\end{tabular}
\end{ruledtabular}
\end{table}

The first regime, shown in Fig.~\ref{fig:density_regimes}(a), is upstream propagation on the subsonic side. For $X_0=20$ and $u_0=-0.75$, the soliton moves from $X_s(0)\simeq 19.99$ to $X_s(t_{\max})\simeq 12.94$, with mean velocity $\langle\dot X_s\rangle\simeq -0.080$. This confirms that upstream motion is possible where $v_0<c_0$ and the soliton relative speed is sufficiently large in magnitude.

The second regime, shown in Fig.~\ref{fig:density_regimes}(b), is finite-depth stalling on the subsonic side. For $X_0=20$ and $u_0=-0.60$, the soliton is initially directed against the flow but does not move appreciably upstream. Instead, it drifts from $X_s(0)\simeq 20.10$ to $X_s(t_{\max})\simeq 24.76$, with $\langle\dot X_s\rangle\simeq 0.053$. Because the horizon is at $x_H\simeq 41$, this case should not be described as a literal horizon collision. It is better interpreted as a finite-depth approach to the branch separatrix: upstream motion is not absolutely forbidden on the subsonic side, but this particular branch does not provide enough negative relative velocity to overcome the local flow.

The third regime, shown in Fig.~\ref{fig:density_regimes}(c), is downstream advection in the supersonic region. For $X_0=70$ and $u_0=-0.60$, the soliton is initialized beyond the acoustic horizon with a negative velocity relative to the fluid. Nevertheless, the extracted center moves from $X_s(0)\simeq 70.35$ to $X_s(t_{\max})\simeq 106.33$, with $\langle\dot X_s\rangle\simeq 0.404$. This is the central numerical result of the current data set. It confirms the branch prediction for this representative supersonic upstream attempt: the defect is carried downstream rather than sustaining a negative laboratory velocity.

Figure~\ref{fig:velocity_bound}(a)--\ref{fig:velocity_bound}(c) compares the smoothed velocity $\dot X_s(t)$ with the local branch edge $v_0-c_0$ evaluated along each representative trajectory. In the subsonic upstream case, Fig.~\ref{fig:velocity_bound}(a), the available velocity interval includes negative values, and the soliton realizes $\dot X_s<0$ for most of the evolution. In the stalling case, Fig.~\ref{fig:velocity_bound}(b), the branch edge remains negative, but the finite-depth soliton drifts weakly downstream. In the supersonic case, Fig.~\ref{fig:velocity_bound}(c), the branch edge is positive, and the extracted soliton velocity remains positive. Small endpoint features in the smoothed derivative are post-processing artifacts caused by differentiating a finite time series and are not used in the physical interpretation.

The depth evolution in Fig.~\ref{fig:depth} shows that the defect is not a rigid particle. Its contrast changes as it interacts with the inhomogeneous flow, and small oscillations are generated by acoustic radiation emitted during the initial adjustment to the background. The important point is that the soliton core remains trackable and coherent over the time interval used to establish the kinematic trend.

\begin{figure*}[tbhp]
\centering
\includegraphics[width=0.49\textwidth]{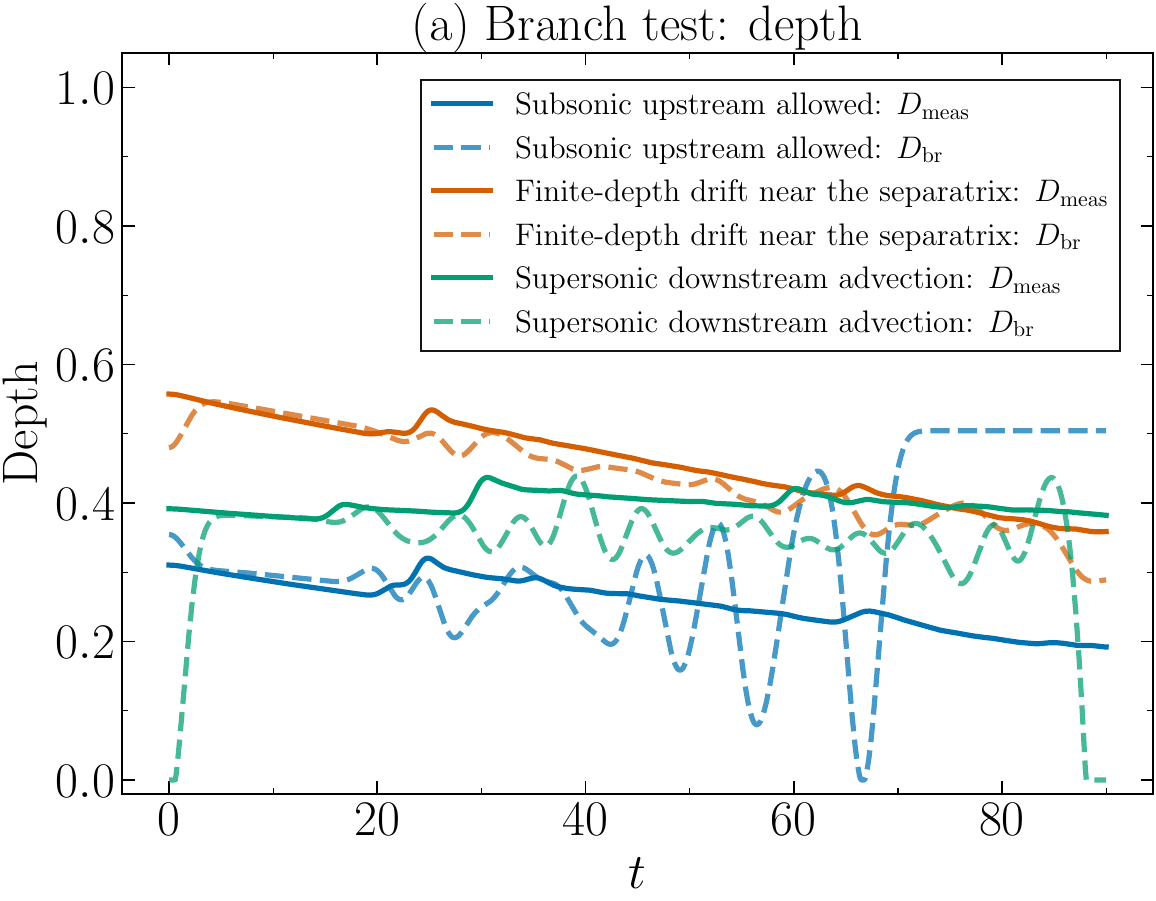}\hfill
\includegraphics[width=0.49\textwidth]{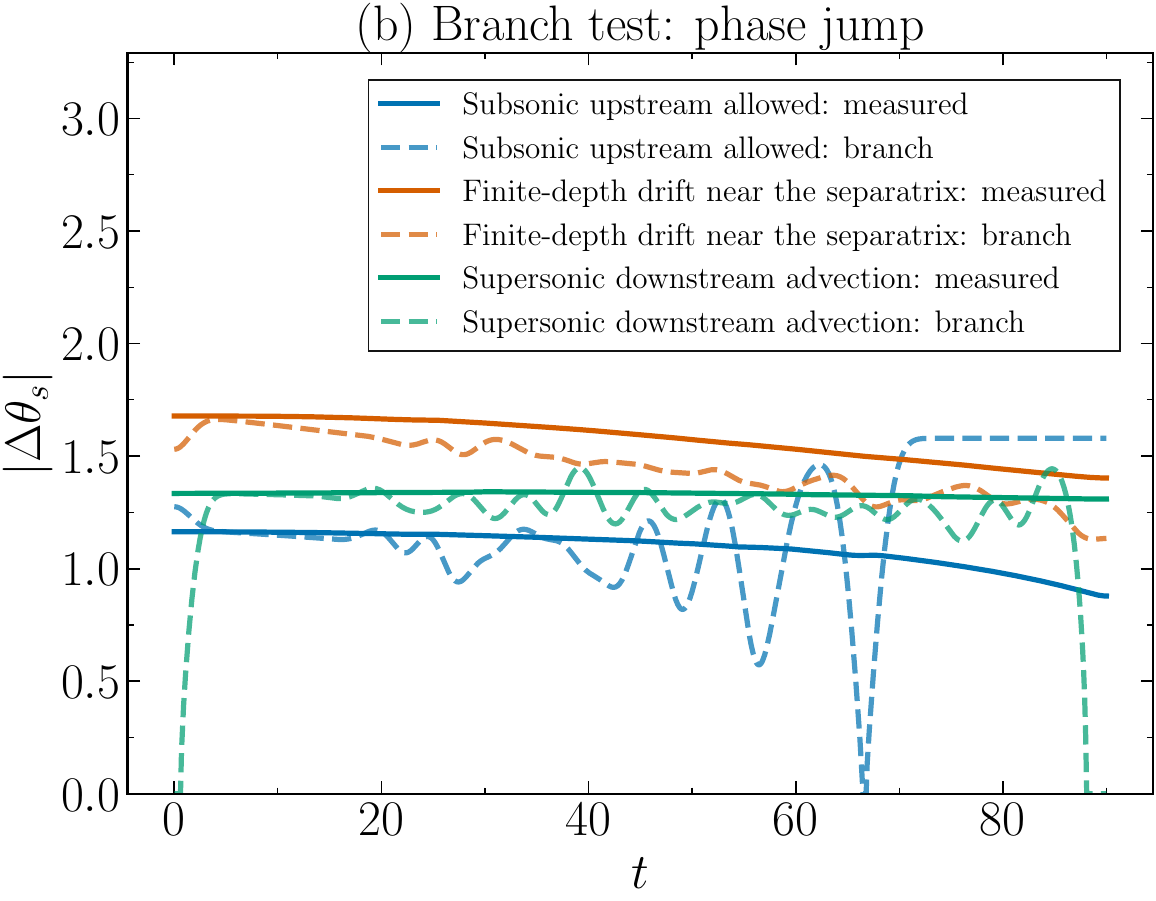}\\[0.5em]
\includegraphics[width=0.49\textwidth]{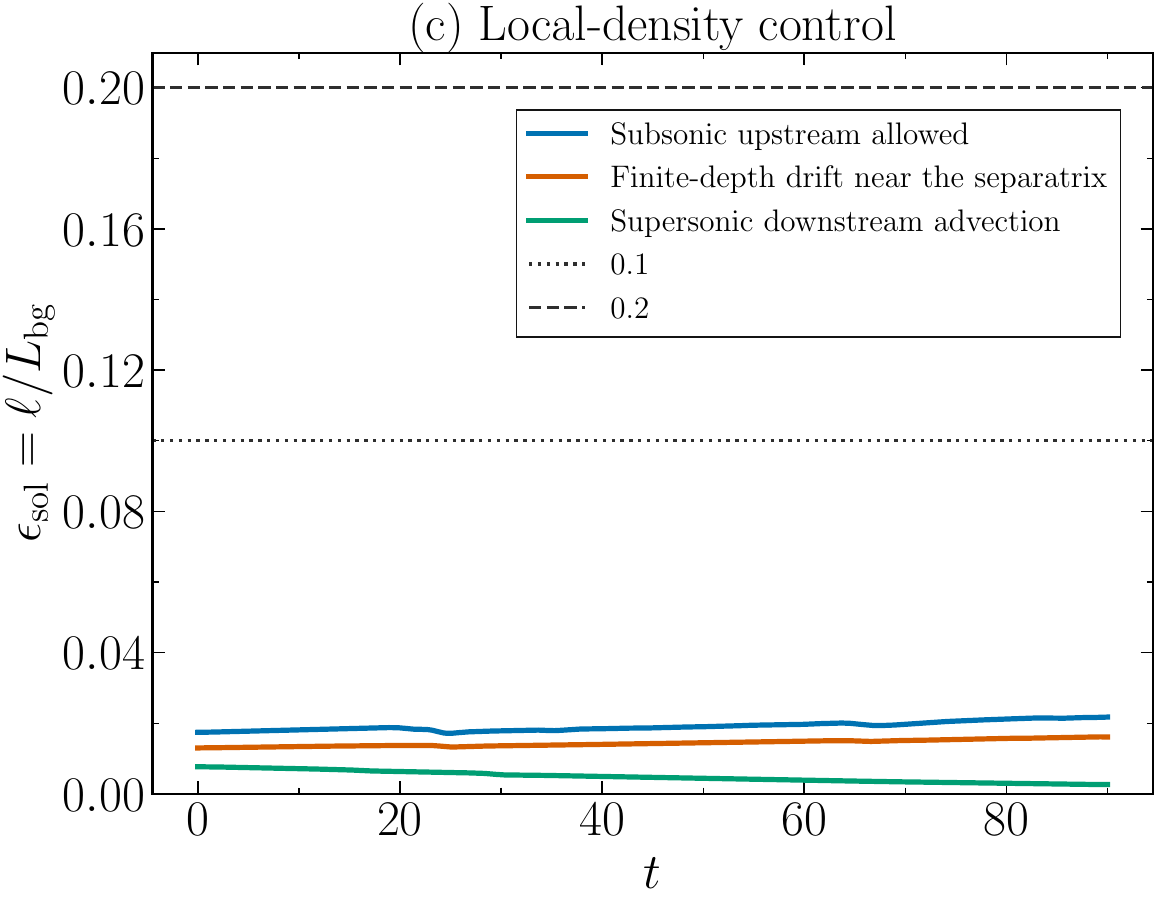}\hfill
\includegraphics[width=0.49\textwidth]{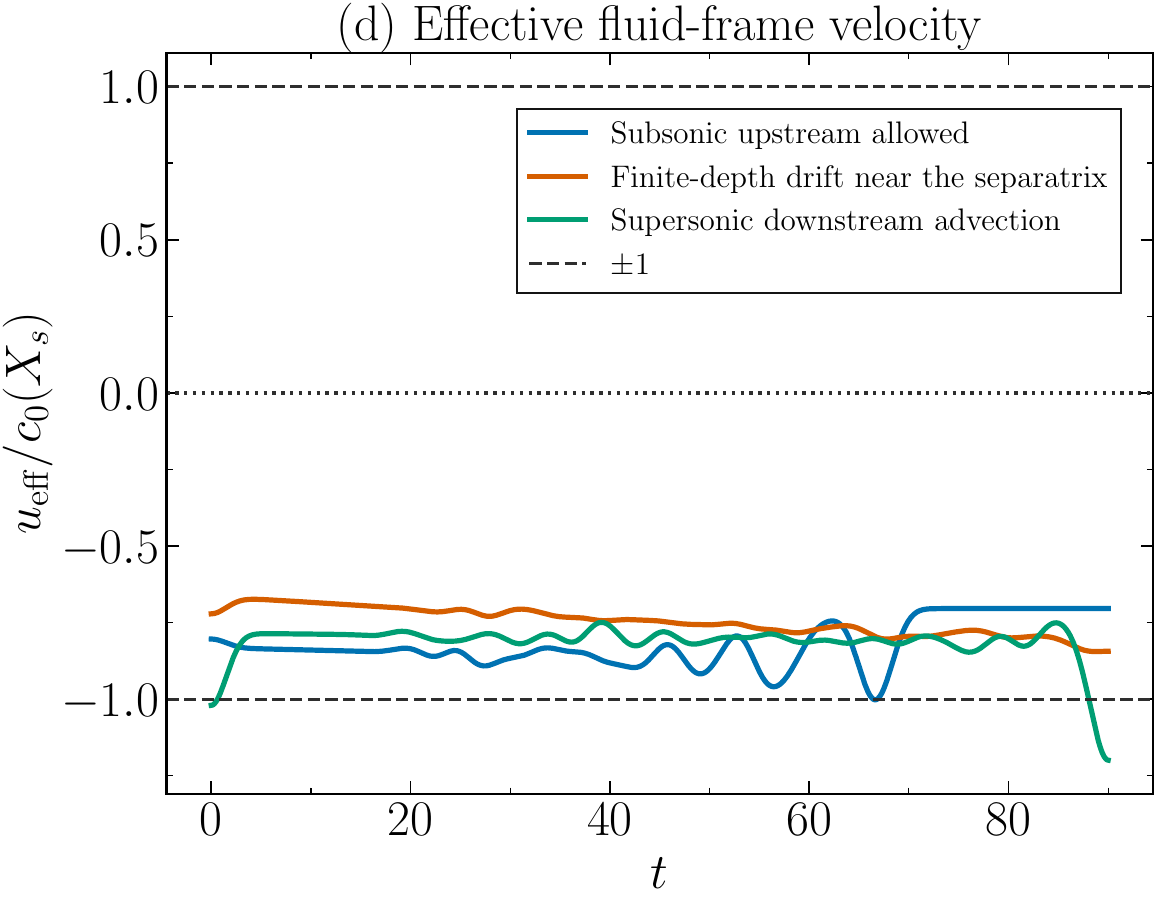}
\caption{Branch-consistency and local-density diagnostics for the three representative cases. (a) Measured density contrast $D_{\rm meas}$ compared with the local homogeneous-branch prediction $D_{\rm branch}=1-\ueff^2/c_0^2$. (b) Measured residual phase jump compared with the branch prediction obtained from $\Delta\theta_{\rm branch}=2\arccos(\ueff/c_0)$, with the same phase-wrapping convention as in the phase-jump diagnostic. (c) Width-based local-density control parameter $\epsilon_{\rm sol}=\ell/L_{\rm bg}$, estimated from the measured depth. (d) Effective fluid-frame velocity ratio $\ueff/c_0[X_s(t)]$; the dashed guide lines mark the regular dark-soliton branch endpoints $\ueff/c_0=\pm1$.}
\label{fig:branch_diagnostics}
\end{figure*}

A nontrivial local branch diagnostic is shown in Fig.~\ref{fig:branch_diagnostics}. The measured depth and residual phase jump are compared with the values inferred from the effective fluid-frame velocity, while the width-based parameter $\epsilon_{\rm sol}=\ell/L_{\rm bg}$ monitors the validity of the local-density picture. The agreement is not expected to be exact, because the profiles are imprinted on an inhomogeneous accelerating background and radiate weak Bogoliubov sound. The useful information is qualitative and diagnostic: the tracked object stays inside the fluid-frame interval $|\ueff|<c_0$ during the relevant evolution, the local-density parameter remains moderate rather than order unity, and the measured phase jump remains finite. These checks support the interpretation of the density minimum as a soliton-like branch excitation rather than as an arbitrary sound depression.

The fixed-$u_0$ scan confirms that the representative cases are not isolated. In the subsonic scan at $X_0=20$, weak upstream attempts are advected toward the horizon, intermediate cases stall, and sufficiently large negative $u_0$ produces upstream motion. In the supersonic scan at $X_0=70$, all tested fixed-$u_0$ cases are advected to larger $x$. That scan establishes the proof-of-principle regimes, while the dimensionless scan below tests the most direct objection: whether a sufficiently shallow, near-sonic soliton could maintain upstream laboratory motion in the supersonic region.

\subsection{Dense upstream-attempt scan in the supersonic region}
\label{subsec:alpha_scan}

The representative supersonic case uses a fixed absolute value of the initial fluid-frame velocity. To verify that the downstream advection is not an artifact of this particular choice, we performed a denser scan at fixed initial position $X_0=70$, where the background Mach number is $M_0\simeq1.25$. The initial velocity was parameterized by Eq.~\eqref{eq:alpha_scan_definition}, with
\begin{equation}
\alpha=0.50,0.60,0.70,0.80,0.90,0.95,0.98.
\label{eq:alpha_values}
\end{equation}
For each value, the initial condition is an upstream attempt in the local fluid frame, $u_0=-\alpha c_0(X_0)$. Values near $\alpha=1$ are the strongest possible upstream attempts that remain on the regular dark-soliton branch; at the same time, they are shallow and broad, so they also test the limit in which local soliton identity becomes most delicate. The endpoint case $\alpha=0.98$ is included deliberately as a marginal near-sonic test, where the density contrast is low and velocity diagnostics based on differentiating a tracked minimum are expected to be more sensitive to tracking and endpoint effects.

\begin{figure*}[tbhp]
\centering
\includegraphics[width=0.95\textwidth]{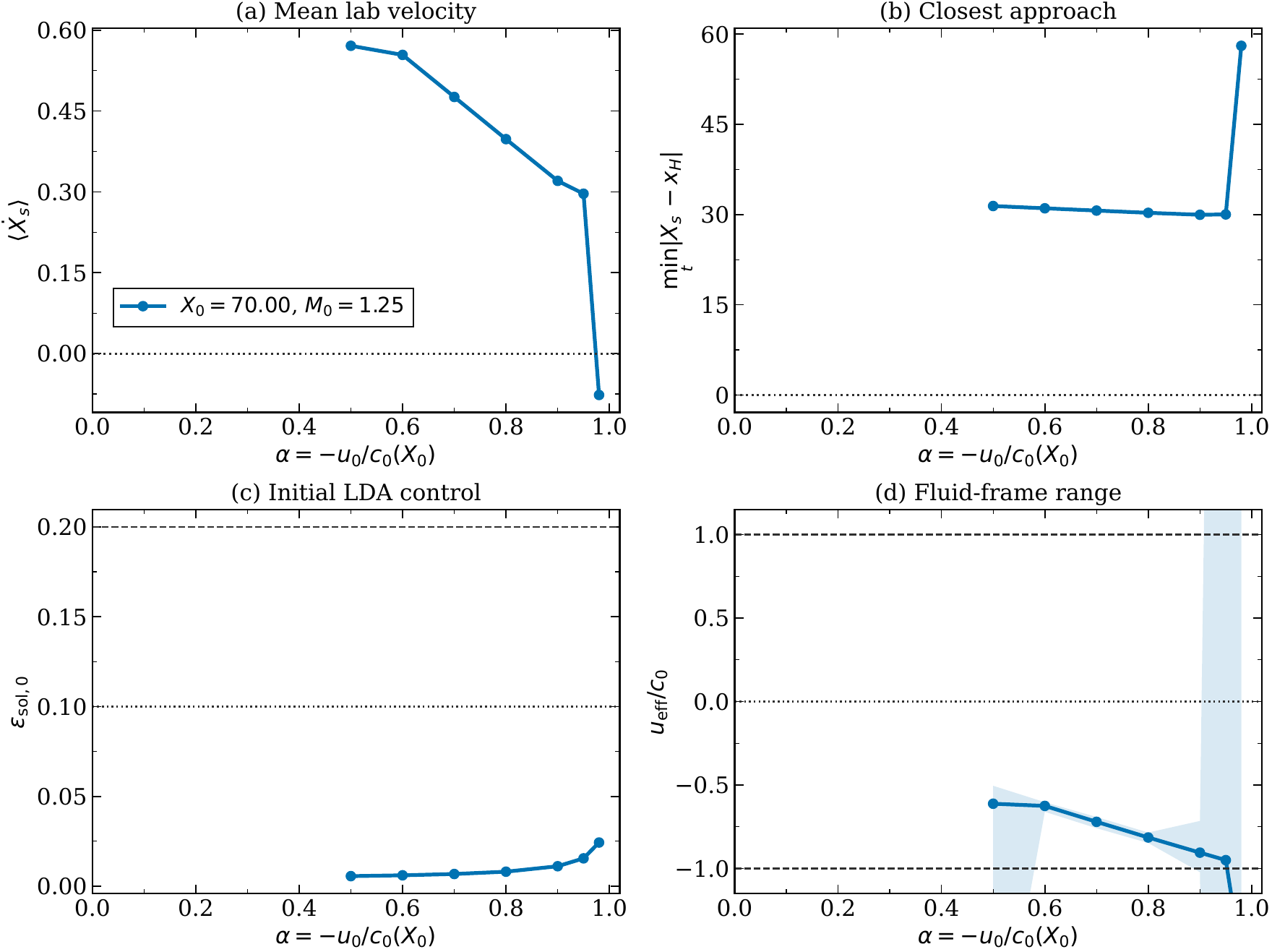}
\caption{Dense scan of upstream attempts launched in the supersonic region at $X_0=70$, where $M_0\simeq1.25$. The initial fluid-frame velocity is parameterized by $\alpha=-u_0/c_0(X_0)$. (a) Mean laboratory velocity of the tracked density minimum. The mean velocity is positive for the robustly tracked cases up to $\alpha=0.95$; the shallow endpoint case $\alpha=0.98$ is shown as a marginal near-sonic diagnostic for which differentiated-minimum velocity extraction is more sensitive. (b) Closest approach to the acoustic horizon. The soliton-like defect does not approach the horizon from the supersonic side; it is advected downstream or becomes tracking-limited in the shallow endpoint regime. (c) Initial local-density control parameter $\epsilon_{{\rm sol},0}=\ell_0/L_{\rm bg}(X_0)$. (d) Range of the effective fluid-frame velocity $\ueff/c_0[X_s(t)]$, showing consistency with the local dark-soliton branch interval $|\ueff|<c_0$ for the robustly tracked cases and increased sensitivity close to the shallow branch endpoint.}
\label{fig:alpha_scan}
\end{figure*}

The result is shown in Fig.~\ref{fig:alpha_scan}. For the robustly tracked cases up to $\alpha=0.95$, the mean laboratory velocity of the density minimum remains positive. The shallowest case, $\alpha=0.98$, lies very close to the low-contrast branch endpoint and is therefore treated as a marginal diagnostic of the shallow limit rather than as a clean quantitative velocity measurement. The trajectory-level information nevertheless shows no sustained upstream branch emerging from the supersonic side. This directly strengthens the branch-blocking interpretation in the supersonic region: the background flow exceeds the maximal upstream velocity supplied by the local soliton branch before a regular dark soliton can maintain upstream laboratory motion.

The same scan provides two internal consistency checks. First, the initial width-based local-density parameter $\epsilon_{{\rm sol},0}=\ell_0/L_{\rm bg}(X_0)$ remains moderate throughout the scan, although it increases for the shallowest solitons. Second, except for the expected increased sensitivity of the marginal endpoint case, the extracted effective fluid-frame velocity $\ueff=\dot X_s-v_0[X_s(t)]$ remains within the local branch interval $|\ueff|<c_0[X_s(t)]$. The branch-depth agreement is only semiquantitative for the shallowest cases, as expected when the soliton becomes broad, weakly contrasted, and more sensitive to radiative background fluctuations. The robust conclusion of the scan is therefore not exact adiabatic integrability, but the absence of a clean sustained upstream dark-soliton branch throughout a dense sequence of upstream attempts in the supersonic region.

\subsection{Numerical robustness of the supersonic case}
\label{subsec:numerical_robustness}

\begin{figure}[tbhp]
\centering
\includegraphics[scale=0.4]{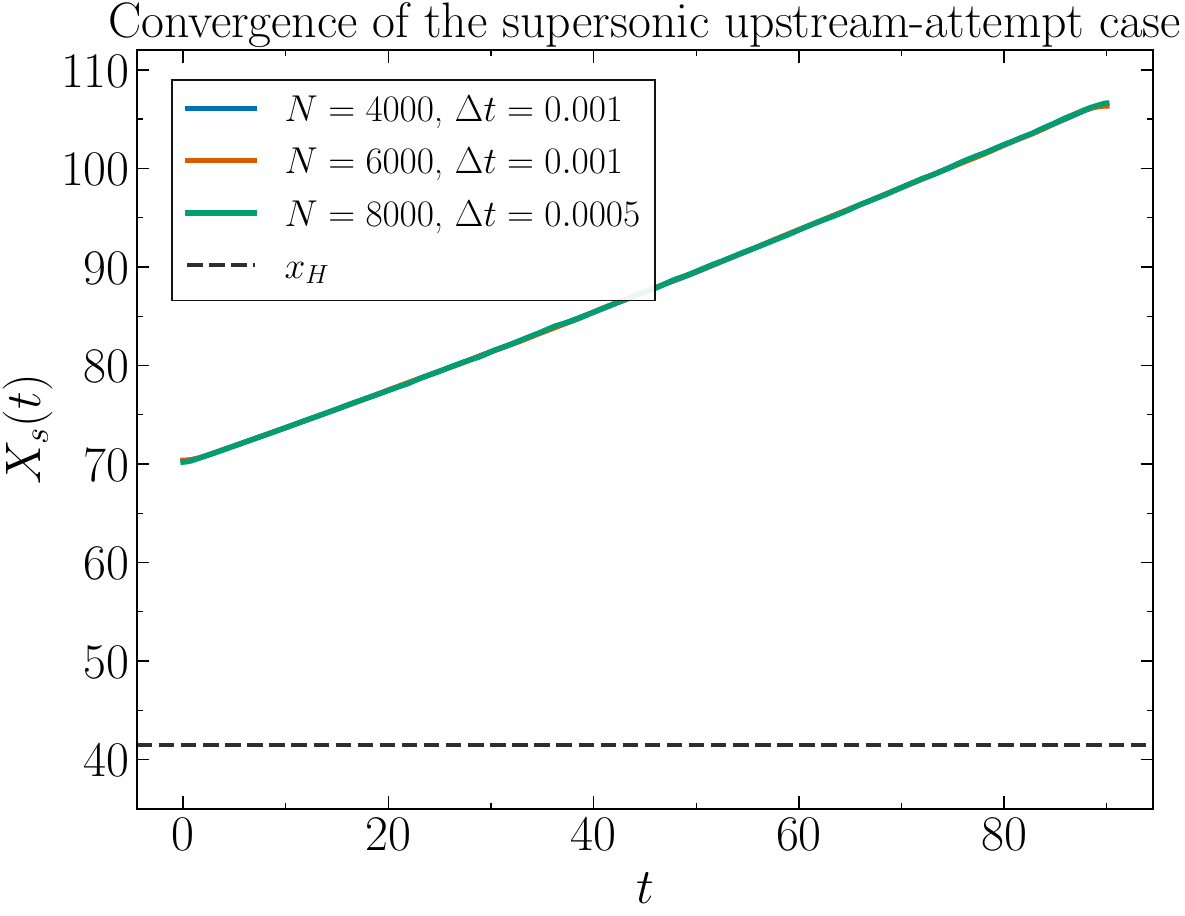}
\caption{Convergence test for the central supersonic upstream-attempt case, $X_0=70$ and $u_0=-0.60$. The soliton trajectory is nearly unchanged when the grid is refined from $N=4000$ to $N=8000$ and the time step is reduced from $10^{-3}$ to $5\times10^{-4}$. The result confirms that the forced downstream advection in the supersonic region is not a discretization artifact.}
\label{fig:convergence}
\end{figure}

The most important numerical test in the current data set is the stability of the central supersonic upstream-attempt case, $X_0=70$ and $u_0=-0.60$, under changes of grid spacing and time step. This is the case in which a soliton-like defect is initialized on the supersonic side with negative velocity relative to the fluid, but is nevertheless advected downstream. We repeated this run for three discretizations:
\begin{equation}
(N,\Delta t)=(4000,10^{-3}),\;
(6000,10^{-3}),\;
(8000,5\times10^{-4}).
\end{equation}
The resulting trajectories are shown in Fig.~\ref{fig:convergence}. The extracted final positions are
\begin{equation}
X_s(t_{\max})=106.38,\; 106.33,\; 106.63,
\end{equation}
and the corresponding mean velocities are
\begin{equation}
\langle\dot X_s\rangle=0.4046,\quad 0.4044,\quad 0.4069.
\end{equation}
The mean velocity therefore changes by less than one percent across the tested discretizations. The downstream advection in the supersonic region is not a grid or time-step artifact.

Additional diagnostics are given in Appendix~\ref{app:additional_diagnostics}. There, we show the fixed-$u_0$ upstream scan, supplementary views of the dense $\alpha$ scan, and an independent phase-jump diagnostic. The latter supports the interpretation that the tracked density depletion retains a soliton-like phase signature during the time interval used to test the branch criterion.

\section{Discussion}
\label{sec:discussion}

The acoustic horizon is defined by the linear Bogoliubov sector, yet the same sonic condition constrains a nonlinear dark soliton. The reason is that the local sound speed enters the two problems in complementary ways. For phonons, $c_0$ is the long-wavelength group velocity in the fluid frame. For a dark soliton, $c_0$ is the endpoint of the nonlinear soliton branch: as the fluid-frame soliton speed approaches $c_0$, the density depletion becomes shallow, the width grows, and the defect merges with the sound continuum. The numerical results show that this branch-existence statement is consistent with the full Gross--Pitaevskii evolution in the representative inhomogeneous open background considered here. Upstream motion is observed on the subsonic side, a finite-depth subsonic attempt stalls, and the robustly tracked soliton-like defects initialized on the supersonic side are advected downstream.

The proposed mechanism is therefore best understood as a nonlinear speed-limit effect, not as a geodesic approximation. The soliton is not assumed to follow null rays of the acoustic metric. Instead, the same local hydrodynamic cone that blocks phonons also bounds the relative speed available to the soliton branch. Once the background flow reaches the sound speed, the maximal upstream dark-soliton speed in the local fluid frame is exhausted. Once $v_0>c_0$, a regular local soliton branch cannot maintain a negative laboratory velocity.

This distinction is important for interpreting the stalling case. The trajectory is not a perfect reflection at $x_H$ and should not be described as a hard-wall collision. In the present run, the soliton starts at $X_0=20$ while the horizon is at $x_H\simeq41$, and the tracked center remains well on the subsonic side. It is therefore a finite-depth soliton whose available negative relative velocity is insufficient to overcome the local convective flow. The density maps show weak acoustic radiation and a slow change of contrast, as expected when a homogeneous-soliton profile is imprinted on an inhomogeneous accelerating background. These effects change the quantitative trajectory, but they do not invalidate the branch picture. The relevant statement is that the upstream branch terminates at the sonic condition; individual solitons may stall, turn, deform, or lose contrast before reaching it.

The local-density approximation also has a controlled weakness. The parameter $\xi/L_{\rm bg}$ is not enough for a grey soliton. The soliton width is $\ell=\xi/\sqrt D$, so the adiabatic parameter $\epsilon_{\rm sol}=\ell/L_{\rm bg}$ grows as the branch approaches the shallow sonic limit. This observation is physically important rather than merely technical: the soliton state that would come closest to overcoming a sonic flow is also the state least sharply localized as a soliton. The representative diagnostics in Fig.~\ref{fig:branch_diagnostics} report $\epsilon_{\rm sol}(t)$ together with the branch variables, and the dense scan in Fig.~\ref{fig:alpha_scan} applies the same logic to $\alpha=-u_0/c_0(X_0)$ values approaching unity. The fact that $\epsilon_{{\rm sol},0}$ remains moderate in that scan strengthens the interpretation, while the increasing branch-depth residual for shallow cases reminds us that the near-sonic endpoint is only semiquantitatively adiabatic.

The one-dimensional geometry has a physical role beyond numerical simplicity. In one dimension, the dark soliton is a robust coherent excitation of the defocusing Gross--Pitaevskii equation. In higher dimensions, planar dark solitons are susceptible to transverse snake instabilities and may decay into vortex pairs or vortex rings. Such physics is interesting, but it would obscure the minimal branch-existence mechanism tested here. The present one-dimensional model is therefore the cleanest setting in which to isolate the nonlinear analogue of acoustic blocking. A natural extension would be to ask whether the blocked soliton branch in higher dimensions converts into vortical excitations rather than simply being advected.

The diagnostic value of the soliton trajectory is also worth emphasizing. A phonon probes the linear causal cone, whereas a dark soliton probes the nonlinear coherent sector of the same condensate. Because dark solitons can be imprinted and monitored in elongated condensates, the proposed effect suggests a possible nonlinear diagnostic of transonic flows complementary to phonon-based probes. Observing upstream soliton motion identifies a region with sufficient subsonic margin; observing finite-depth stalling indicates approach to the branch edge for that particular soliton; and observing unavoidable downstream advection identifies the supersonic side. In guided condensates or atomtronic channels, this suggests a practical nonlinear probe of transonic flow profiles. The same mechanism could also be interpreted as a directional filter for coherent nonlinear defects: a transonic flow can transmit or block soliton-like excitations depending on where they are launched and how close their fluid-frame velocity is to the local sound-speed bound.

Several limitations should be kept explicit. A quantitative comparison with a specific experiment would require modeling the preparation of the transonic background, the quasi-one-dimensional coupling, transverse confinement, finite temperature, and imaging protocol; these ingredients are beyond the scope of the present proof-of-principle study. First, the present work uses a prescribed stationary background constructed by inverting the Bernoulli relation. This is appropriate for isolating the branch mechanism, but it does not address how such a profile is produced experimentally in a particular trap or atomtronic device. Second, the scans are still not a full phase diagram. The fixed-$u_0$ scans demonstrate the three regimes, and the dense $\alpha$ scan tests the strongest upstream attempts at one supersonic launch point, but we have not yet varied the horizon gradient or performed an equally dense set of launch positions $X_0\simeq x_H\pm\delta$. Third, the trajectory is extracted from the density minimum and then smoothed for velocity diagnostics; this is suitable while the defect remains soliton-like, but it would become ambiguous after strong radiation or soliton decay. The phase-jump diagnostic in Appendix~\ref{app:phase_jump} is included precisely to support the soliton-like interpretation over the time interval used here. Fourth, the present post-processing does not yet provide a full simultaneous fit of density profile, phase jump, and effective velocity to the homogeneous soliton relations; Eqs.~\eqref{eq:D_branch} and \eqref{eq:theta_branch} define the appropriate next diagnostic.

The most direct future extension is a two-parameter scan over initial grayness and horizon gradient. The present dense scan varies the grayness parameter at a fixed supersonic launch point; the next step is to repeat the same branch diagnostics for launches close to the horizon and for different values of the horizon gradient. The kinematic argument predicts that deeper solitons, which have smaller $|u|$, should stall farther from the horizon, whereas shallow solitons should approach the sonic point more closely before losing contrast. A second extension is to quantify emitted Bogoliubov radiation and separate the coherent soliton momentum from the radiative momentum. A third is to replace the designed stationary background with a dynamically generated transonic flow in a specific quasi-one-dimensional trap. These steps would turn the current proof of principle into a quantitative protocol for nonlinear horizon spectroscopy.

\section{Conclusion}
\label{sec:conclusion}

We have formulated and numerically tested the branch-blocking problem for a one-dimensional dark soliton in a transonic Gross--Pitaevskii background. Starting from the stationary condensate hydrodynamics, we constructed transonic backgrounds, identified the acoustic horizon, reviewed the intrinsic dark-soliton velocity bound, and combined these ingredients into the local branch interval
\begin{equation}
v_0(X_s)-c_0(X_s)<\dot X_s<v_0(X_s)+c_0(X_s)
\end{equation}
at leading local-density order. For a right-moving flow, a regular local dark-soliton branch can sustain upstream laboratory motion only on the subsonic side. The sonic point is therefore the limiting separatrix of the regular soliton branch, while finite-depth solitons may stall, turn, deform, radiate, or lose contrast before reaching it.

The numerical simulations solve the full complex Gross--Pitaevskii equation in an open, nonperiodic domain and extract all observables from the evolved field. The stationary background is validated by a residual $\epsilon_R\simeq1.6\times10^{-4}$ and remains stable throughout the evolution. The resulting trajectories exhibit three representative regimes: upstream propagation on the subsonic side, finite-depth stalling on the subsonic side, and forced downstream advection in the supersonic region. The last regime provides the central evidence in the present data set for nonlinear branch blocking: a dark-soliton-like defect initialized with a negative velocity relative to the fluid does not sustain upstream laboratory motion where $v_0>c_0$. This conclusion is strengthened by a dense scan at $X_0=70$ over $\alpha=-u_0/c_0(X_0)$. The mean laboratory velocity remains positive for the robustly tracked cases up to $\alpha=0.95$, while the endpoint case $\alpha=0.98$ exhibits the expected tracking sensitivity of a very shallow, broad defect close to the branch endpoint. Across the scan, no clean sustained upstream dark-soliton branch is observed on the supersonic side.

The main contribution is thus a clean identification of a nonlinear horizon-related branch constraint that does not require assigning a metric trajectory to the soliton. The dark soliton inherits the sonic restriction through its own branch-existence condition. This makes it a complementary probe of transonic condensate flows, sensitive not only to the linear phonon cone but also to the transport of coherent nonlinear defects. The framework opens a route to quantitative studies of nonlinear horizon scattering, radiation emission, phase-jump evolution, and the breakdown or conversion of soliton identity near transonic transitions.

\appendix

\section{Additional numerical diagnostics}
\label{app:additional_diagnostics}

This appendix collects diagnostics that support the interpretation of the main text. They are not needed to identify the three regimes in Figs.~\ref{fig:density_regimes} and \ref{fig:trajectories}, but they address three natural numerical questions: whether the representative cases are part of a systematic fixed-$u_0$ scan, how the dense dimensionless scan behaves at the trajectory level, and whether the tracked density minimum retains a soliton-like phase signature.

\subsection{Full upstream scan}
\label{app:full_scan}

\begin{figure}[tbhp]
\centering
\includegraphics[scale=0.43]{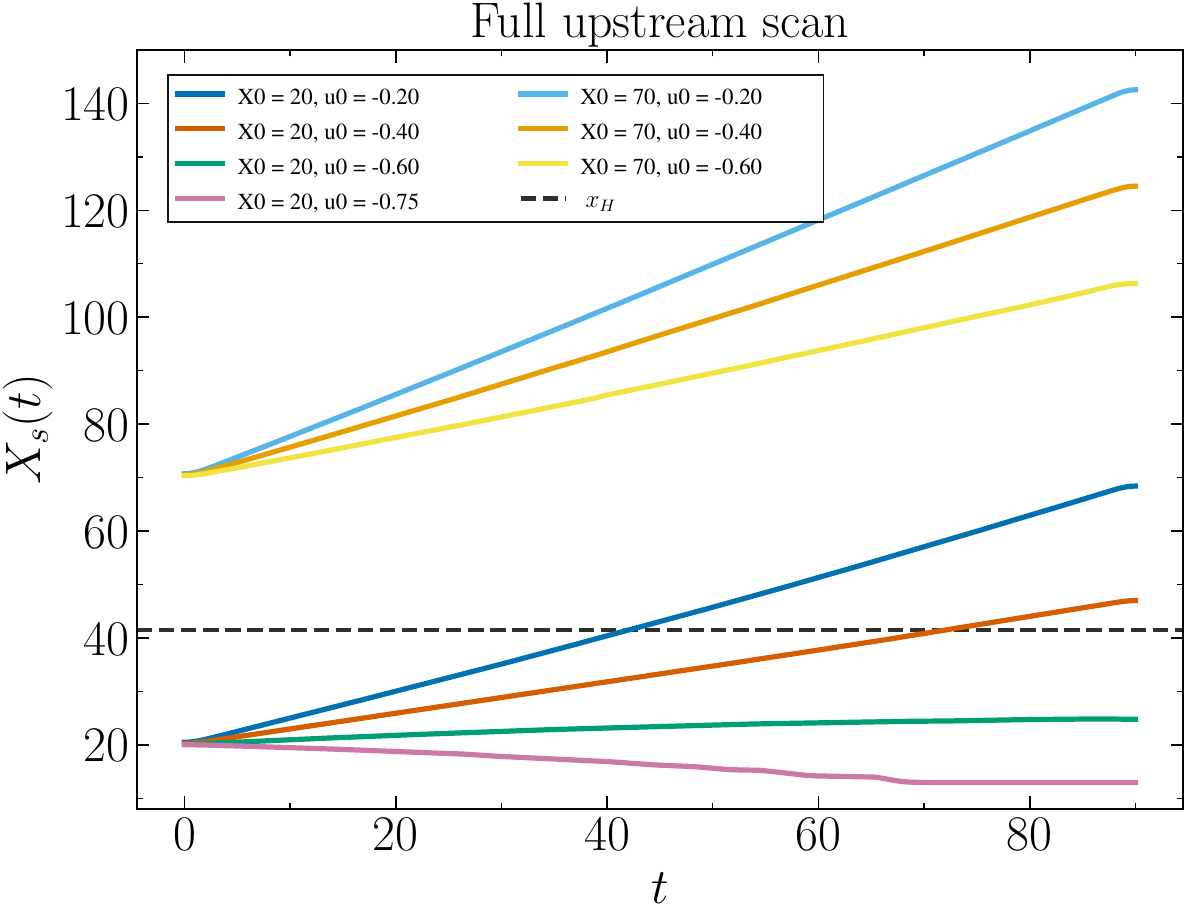}
\caption{Full upstream scan. The subsonic cases initialized at $X_0=20$ show the transition from downstream advection to stalling and upstream propagation as the relative velocity becomes more negative. The supersonic cases initialized at $X_0=70$ are all advected downstream in the tested set. The dashed horizontal line marks $x_H$.}
\label{fig:all_scan}
\end{figure}

Figure~\ref{fig:all_scan} shows the fixed-$u_0$ upstream scan. On the subsonic side, with $X_0=20$, the trajectories interpolate between downstream advection, finite-depth stalling, and upstream propagation as $u_0$ becomes more negative. On the supersonic side, with $X_0=70$, all tested trajectories move toward increasing $x$, despite the negative initial velocity relative to the fluid. This confirms that the three representative cases shown in the main text were not selected in isolation, but summarize a broader family of initial conditions. The targeted dimensionless scan in Fig.~\ref{fig:alpha_scan} and Fig.~\ref{fig:alpha_scan_trajectories} complements this fixed-$u_0$ scan by sampling $\alpha=-u_0/c_0(X_0)$ close to unity.

\subsection{Supplementary views of the dense $\alpha$ scan}
\label{app:alpha_scan_supplement}

\begin{figure}[tbhp]
\centering
\includegraphics[scale=0.43]{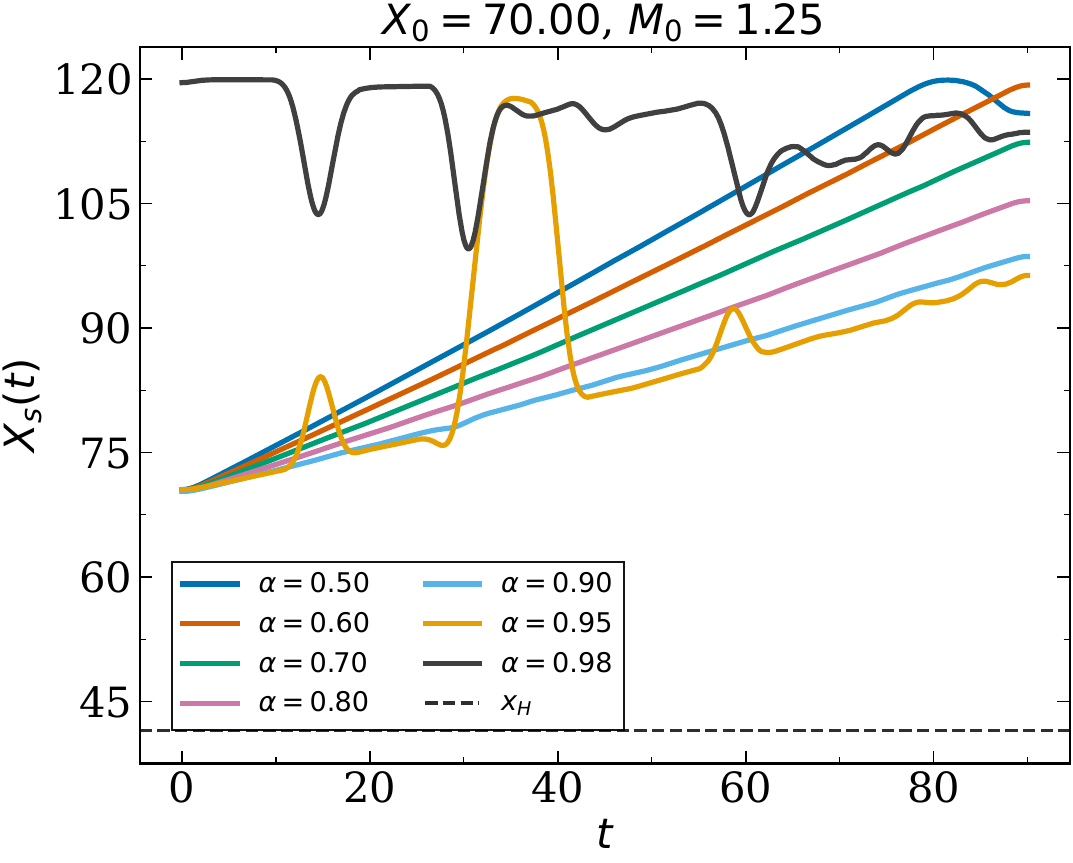}
\caption{Trajectories in the dense dimensionless upstream-attempt scan at fixed supersonic launch point $X_0=70$. The initial velocity is $u_0=-\alpha c_0(X_0)$, with $\alpha$ ranging from $0.50$ to $0.98$. The tracked defects remain on the downstream side of the horizon and do not establish a sustained upstream branch; the shallowest endpoint case is the most tracking-sensitive, as expected near the low-contrast branch limit. The dashed line marks $x_H$.}
\label{fig:alpha_scan_trajectories}
\end{figure}

The summary panel in Fig.~\ref{fig:alpha_scan} compresses the dense scan into four scalar diagnostics. Figure~\ref{fig:alpha_scan_trajectories} shows the corresponding trajectories. The tracked defects launched at $X_0=70$ remain on the downstream side of the horizon and show no sustained upstream branch. The shallowest cases, $\alpha=0.95$ and $\alpha=0.98$, are especially relevant: they are the closest tested points to the local sound-speed endpoint of the dark-soliton branch. The $\alpha=0.98$ trajectory should be read as a marginal low-contrast endpoint test, where tracking sensitivity is expected to be strongest.

\begin{figure}[tbhp]
\centering
\includegraphics[scale=0.43]{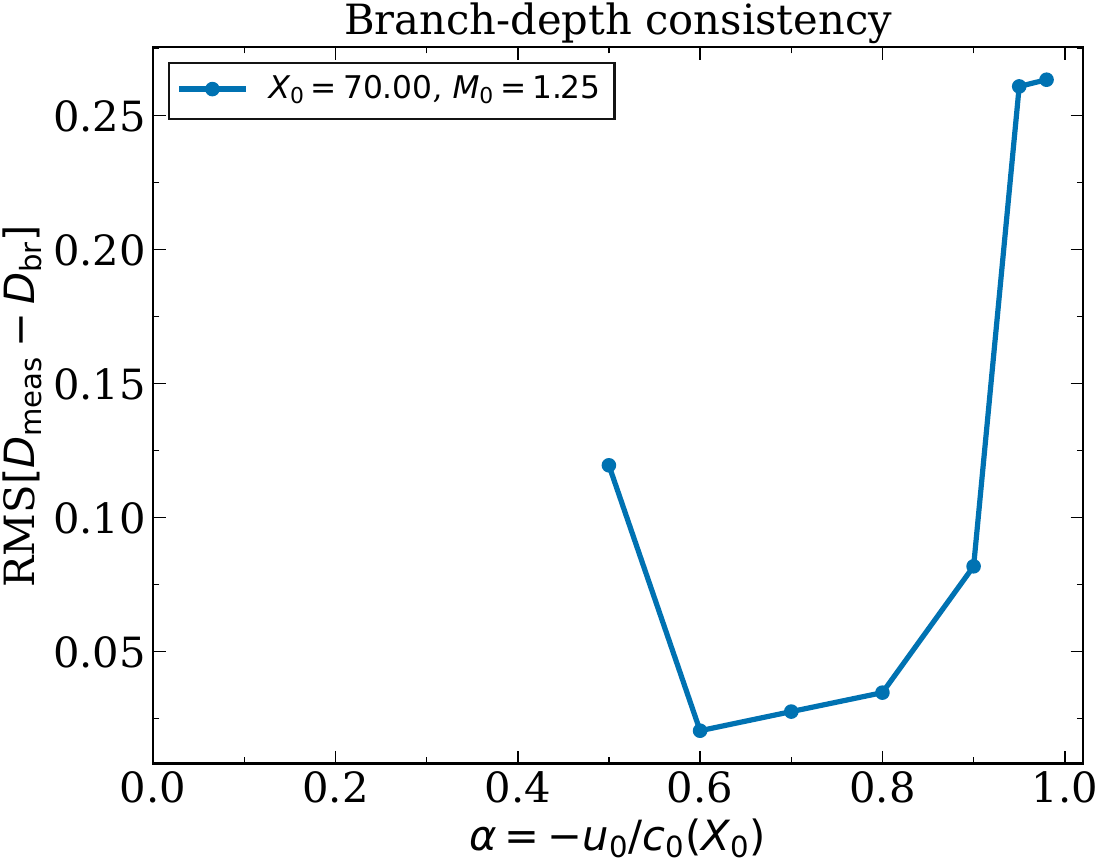}
\caption{Branch-depth consistency in the dense dimensionless scan. The residual $\mathrm{RMS}[D_{\rm meas}-D_{\rm branch}]$ is semiquantitative and increases for shallow near-sonic solitons, as expected when the defect becomes broad and weakly contrasted.}
\label{fig:alpha_scan_depth_rms}
\end{figure}

Figure~\ref{fig:alpha_scan_depth_rms} reports a compact branch-depth residual for the same scan. The residual grows for shallow solitons, which is expected because the density contrast decreases and the soliton width increases as $\alpha\to1$. This trend should not be interpreted as a failure of the blocking result. It indicates that the shallow near-sonic edge is the least ideal regime for a quantitative homogeneous-branch fit, while the branch-blocking interpretation remains robust for the cleanly tracked cases.

\subsection{Phase-jump diagnostic}
\label{app:phase_jump}

The trajectory is extracted from the density minimum, but a dark soliton is also characterized by a phase jump across its core. To check that the tracked density depletion remains soliton-like, we subtract the stationary background phase $\theta_0(x)$ and estimate the residual phase jump across the core,
\begin{equation}
\Delta\theta_s(t)=
\left[
\theta(x_R,t)-\theta_0(x_R)
\right]
-
\left[
\theta(x_L,t)-\theta_0(x_L)
\right],
\label{eq:phase_jump_diag}
\end{equation}
where $x_L$ and $x_R$ are points or small averaging windows on the left and right sides of the tracked core. The plotted quantity is the wrapped absolute value $|\Delta\theta_s(t)|$.

\begin{figure}[tbhp]
\centering
\includegraphics[scale=0.43]{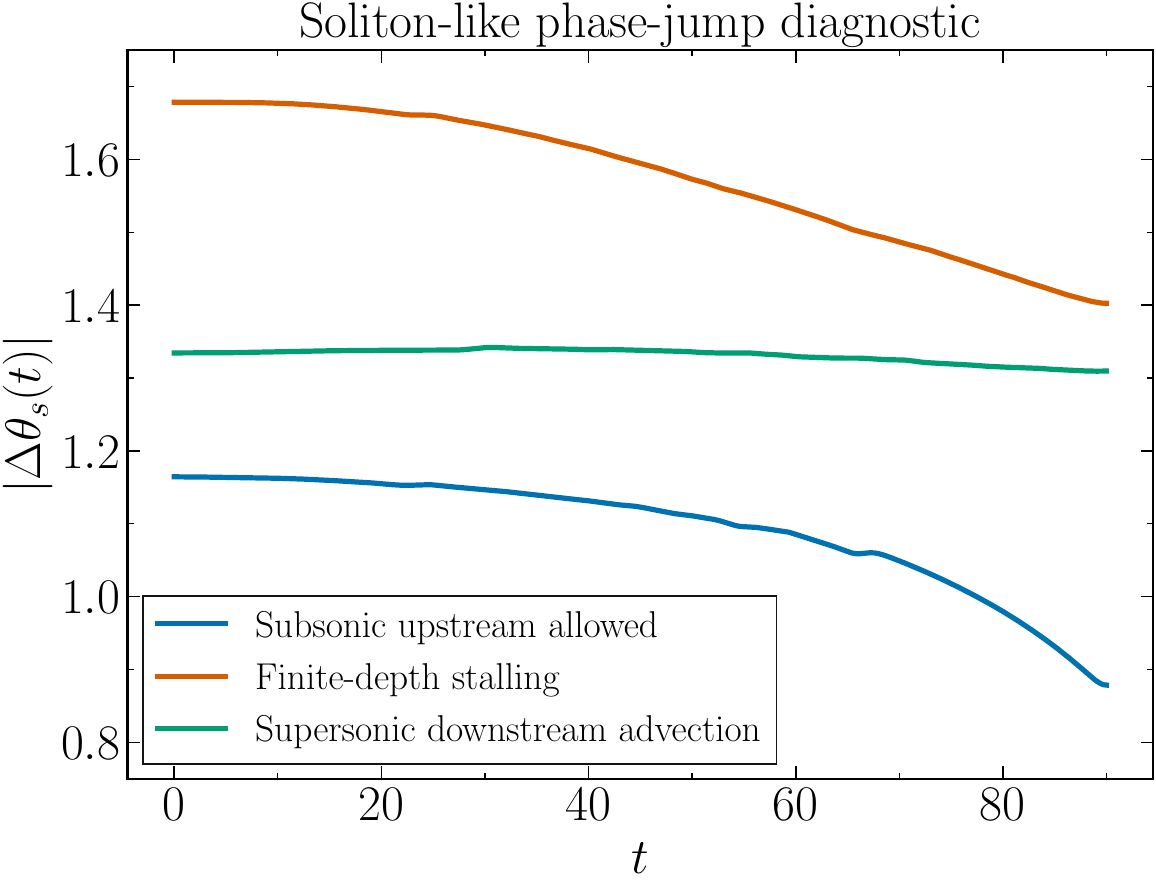}
\caption{Phase-jump diagnostic for the three representative regimes. After subtracting the stationary background phase, the residual phase jump across the tracked density depletion remains finite. This confirms that the tracked objects retain a soliton-like phase signature during the relevant part of the evolution.}
\label{fig:phase_jump_rep}
\end{figure}

Figure~\ref{fig:phase_jump_rep} shows the phase-jump diagnostic for the three representative regimes. The phase jump remains finite throughout the evolution. For the three cases in Table~\ref{tab:regimes}, the mean values are approximately
\begin{equation}
\langle|\Delta\theta_s|\rangle\simeq
1.10,\quad 1.58,\quad 1.33,
\end{equation}
for the subsonic upstream, subsonic finite-depth stalling, and supersonic downstream-advection regimes, respectively. In the central supersonic case, the phase jump changes only from approximately $1.334$ to $1.310$ during the simulation. This supports the interpretation that the object advected downstream in the supersonic region is still a coherent soliton-like defect over the interval used to test the branch criterion. The diagnostic is not meant to prove exact integrable-soliton behavior in an inhomogeneous background; it verifies that the tracked density minimum retains the phase-defect character expected of a dark soliton.

\begin{acknowledgments}
This work was supported by CAPES (Finance Code 001), CNPq (Grant PQ/306308/2022-3), and FAPEMA (Grants UNIVERSAL-06395/22).
\end{acknowledgments}

%

%\bibliographystyle{apsrev4-2}
%\bibliography{References}

\end{document}